\documentclass{siamonline250211}

\usepackage{amsmath,amssymb,amsfonts}
\usepackage{graphicx}
\usepackage{hyperref}
\hypersetup{
    colorlinks=true,
    linkcolor=red,
    filecolor=red,     
    urlcolor=red,
    citecolor=red
}
\usepackage{url}
\usepackage{algorithm}
\usepackage{algpseudocodex}
\usepackage{multirow}
\usepackage{booktabs}
\usepackage{enumitem}
\usepackage{mathtools}
\usepackage{caption}
\usepackage{wrapfig}

\newtheorem{condition}{Condition}[section]

\usepackage{accents} 

\newcommand{\D}{\mathrm{d}}

\renewcommand{\tt}{{\widetilde{t}}}

\newcommand{\ba}{{\boldsymbol{a}}}

\newcommand{\bb}{{\boldsymbol{b}}}

\newcommand{\bE}{{\boldsymbol{E}}}

\newcommand{\bg}{{\boldsymbol{g}}}

\newcommand{\bK}{{\boldsymbol{K}}}

\newcommand{\bn}{{\boldsymbol{n}}}

\newcommand{\bx}{{\boldsymbol{x}}}
\newcommand{\bX}{{\boldsymbol{X}}}
\newcommand{\by}{{\boldsymbol{y}}}

\newcommand{\bz}{{\boldsymbol{z}}}
\newcommand{\bZ}{{\boldsymbol{Z}}}

\newcommand{\bDelta}{{\boldsymbol{\Delta}}}

\newcommand{\bvarepsilon}{{\boldsymbol{\varepsilon}}}
\newcommand{\bzeta}{{\boldsymbol{\zeta}}}
\newcommand{\bEta}{{\boldsymbol{\eta}}}
\newcommand{\btheta}{{\boldsymbol{\theta}}}

\newcommand{\bmu}{{\boldsymbol{\mu}}}

\newcommand{\bxi}{{\boldsymbol{\xi}}}

\newcommand{\btau}{{\boldsymbol{\tau}}}

\newcommand{\bPhi}{{\boldsymbol{\Phi}}}
\newcommand{\bchi}{{\boldsymbol{\chi}}}

\newcommand{\hbmu}{{\widehat{\boldsymbol{\mu}}}}

\newcommand{\hbtau}{{\widehat{\boldsymbol{\tau}}}}

\newcommand{\tblambda}{{\widetilde{\boldsymbol{\lambda}}}}

\newcommand{\hK}{{\widehat{K}}}

\newcommand{\hmu}{{\widehat{\mu}}}

\newcommand{\htau}{{\widehat{\tau}}}

\renewcommand{\sf}{{\starhat{f}}}

\newcommand{\ma}{{\mathsf{a}}}
\newcommand{\mA}{{\mathsf{A}}}
\newcommand{\mb}{{\mathsf{b}}}
\newcommand{\mB}{{\mathsf{B}}}

\newcommand{\mD}{{\mathsf{D}}}

\newcommand{\mE}{{\mathsf{E}}}

\newcommand{\mF}{{\mathsf{F}}}

\newcommand{\mG}{{\mathsf{G}}}

\newcommand{\mH}{{\mathsf{H}}}
\newcommand{\mi}{{\mathsf{i}}}
\newcommand{\mI}{{\mathsf{I}}}

\newcommand{\mJ}{{\mathsf{J}}}

\newcommand{\mK}{{\mathsf{K}}}

\newcommand{\mL}{{\mathsf{L}}}

\newcommand{\mM}{{\mathsf{M}}}

\newcommand{\mR}{{\mathsf{R}}}

\newcommand{\mS}{{\mathsf{S}}}

\newcommand{\mV}{{\mathsf{V}}}

\newcommand{\mx}{{\mathsf{x}}}
\newcommand{\mX}{{\mathsf{X}}}

\newcommand{\mPi}{{\mathsf{\Pi}}}

\newcommand{\mSigma}{{\mathsf{\Sigma}}}

\newcommand{\tmK}{{\widetilde{\mathsf{K}}}}

\newcommand{\tmLambda}{{\widetilde{\mathsf{\Lambda}}}}

\newcommand{\calK}{{\mathcal{K}}}

\newcommand{\calL}{{\mathcal{L}}}

\newcommand{\calN}{{\mathcal{N}}}

\newcommand{\calO}{{\mathcal{O}}}

\newcommand{\calU}{{\mathcal{U}}}

\newcommand{\calX}{{\mathcal{X}}}

\newcommand{\hcalK}{{\widehat{\mathcal{K}}}}

\newcommand{\bbC}{{\mathbb{C}}}

\newcommand{\bbE}{{\mathbb{E}}}

\newcommand{\bbN}{{\mathbb{N}}}

\newcommand{\bbR}{{\mathbb{R}}}

\newcommand{\dell}{{\dot{\ell}}}

\newcommand{\llvert}{\left\lvert}
\newcommand{\rrvert}{\right\rvert}

\newcommand{\bzero}{{\boldsymbol{0}}}
\newcommand{\bone}{{\boldsymbol{1}}}

\newcommand{\Cov}{\mathrm{Cov}}

\newcommand{\qtq}[1]{\quad\mathrm{#1}\quad}

\newcommand{\diag}{\mathrm{diag}}

\newcommand{\starhat}[1]{\accentset{\star}{#1}}

\newcommand\ringring[1]{%
  {
   \mathop{\kern0pt #1}\limits^{
     \vbox to-1.85ex{
       \kern-2ex 
       \hbox to 0pt{\hss\normalfont\kern.1em \r{}\kern-.45em \r{}\hss}%
       \vss 
     }
   }
  }
}

\usepackage{listings}
\definecolor{darkgreen}{rgb}{0,0.6,0}
\lstdefinestyle{Python}{
    showstringspaces=false,
    language        = Python,
    basicstyle      = \small\ttfamily,
    morekeywords = {as},
    keywordstyle    = \color{blue},
    stringstyle     = \color{purple},
    commentstyle    = \color{darkgreen}\ttfamily,
    breaklines = true,
	postbreak=\text{$\hookrightarrow$\space},
	alsoletter = {>,.} ,
    morekeywords = [2]{>>>,...},
    keywordstyle = [2]\color{cyan}\bfseries}

\newcommand{\patchoverfullhbox}{\\}

\title{Fast multitask Gaussian processes, with application to surrogate modeling of the quark-gluon plasma}

\author{Aleksei G. Sorokin\thanks{Department of Statistics, University of Chicago (sorokin@uchicago.edu).}
\and Pieterjan Robbe\thanks{Sandia National Laboratories, Livermore.}
\and Yen-Chun Liu\thanks{Department of Statistical Science, Duke University.}
\and Simon Mak\footnotemark[3]
\and Fred~J.~Hickernell\thanks{Department of Applied Mathematics, Illinois Institute of Technology.}
}

\begin{document}

\maketitle

\begin{abstract}
    Gaussian processes (GPs) are broadly used for the surrogate modeling of computer experiments with reliable uncertainty quantification. Our motivating application comes from the study of the quark-gluon plasma (QGP), an extreme state of nuclear matter that filled the universe shortly after the Big Bang. To reliably infer properties of the QGP, multiple surrogate models need to be constructed for related particle collision simulation systems (i.e., multiple ``tasks'').  While there is a body of literature on multitask GPs (MTGPs), such models can be computationally expensive with large datasets: they require $\mathcal{O}(N^3)$ work and $\mathcal{O}(N^2)$ storage for model training, where $N$ is the total number of samples over all tasks. To address this, we propose a new fast MTGP approach, which pairs structured low-discrepancy design points, such as Sobol' points, with special kernel forms for efficient and exact model fitting. This pairing of a kernel and potentially different design points of different sizes across tasks provides a structured Gram matrix, e.g., a circulant block matrix, which we exploit via a novel algorithm for efficient Gram matrix inversion and determinant computation. Our algorithm reduces training costs to $\calO(N \log N)$ work and $\calO(N)$ storage in the case of equal sample sizes for each tasks. In the worst case of severely unbalanced sample sizes across tasks, our algorithm may require up to $\calO(N^2)$ work and storage, with continuous interpolation between these best and worst cases depending on the balance of sample sizes across tasks. An open-source Python implementation is made available in the \texttt{FastGPs} package (\url{https://alegresor.github.io/fastgps/}). We demonstrate the effectiveness of our fast MTGP approach on a range of simulation experiments and on our motivating QGP application.
\end{abstract}



\section{Introduction}\label{sec:intro}


With advances in scientific computing, complex physical phenomena can now be reliably modeled by sophisticated computer simulations. Such ``computer experiments'' \cite{gramacy2020surrogates} are broadly used in modern scientific and engineering problems, e.g., the design of aircraft engines \cite{miller2024expected} and personalized aortic valve replacements \cite{chen2021function}. Our motivating application is the study of the quark-gluon plasma (QGP; \cite{arslandok2023hot}), an exotic state of matter that filled the universe shortly after the Big Bang. To study this plasma, one requires a virtual simulator that mimics the fine-scale evolution of particles colliding at near-light speeds. Such a simulator is highly complex and time-consuming: each run can take thousands of CPU hours to perform \cite{ehlers2025bayesian}. To ensure reliable findings, one further needs to consider \textit{multiple} virtual simulators that model the particle collision in different ways. Given a limited computational budget, the exploration of these costly simulators over a large input space is a highly challenging task. We propose and demonstrate a method that provides a combined surrogate of the multiple virtual simulators with greater speed and accuracy than existing methods.

A popular solution is to use a surrogate model (or an emulator) to efficiently predict the output $f(\bx)$ of the simulator given input variables $\bx$, where $\bx$ lies in the input domain $\calX \subseteq \mathbb{R}^d$. To train such a model, one first generates training data from the simulator $f$ at $n$ designed input points $\mX \subseteq \calX$, then uses the trained model to predict $f$ at a new untested input point. Gaussian processes (GPs; \cite{gramacy2020surrogates,rasmussen.gp4ml}) are broadly used for surrogate modeling, primarily due to its closed-form predictive distribution, which allows for reliable uncertainty quantification. In the case where multiple simulator models need to be emulated (as in our motivating QGP application), a multitask GP (MTGP; \cite{bonilla2007multi}) can be used. An MTGP is a GP model for predicting $T>1$ related black-box functions (i.e., ``tasks''). Such a model is trained by first generating training data on each task $t$ at $n_t$ design points $\mX_t \subseteq \calX$, then using this data to learn similar features between tasks for joint prediction of the $T$ functions. MTGPs have been broadly explored in the machine learning and surrogate modeling literature \cite{williams2008multi,swersky2013multi,ji2021graphical,ji2022multi}.

Despite their popularity, GPs (and MTGPs) face a key computational challenge given large training datasets, i.e., for large sample sizes $n$. For model training, one needs to evaluate the matrix inverse and determinant of an $n \times n$ Gram matrix, which requires $\calO(n^3)$ computation and $\calO(n^2)$ storage. There is a broad literature on tackling this bottleneck, including preconditioned conjugate gradients \cite{gardner.gpytorch_GPU_conjugate_gradient}, covariance tapering \cite{kaufman2008covariance,kaufman2011efficient}, random Fourier features \cite{rahimi2007random,li2024trigonometric}, Vecchia approximations \cite{scaledvecchia,sauer2022vecchiaapproximated}, and inducing points \cite{snelson.inducing_points_GP,li2025prospar}. The same computational challenge is faced by MTGPs, where model training requires the matrix inverse and determinant of an $N \times N$ Gram matrix, where $N = \sum_{t=1}^T n_t$ is the total sample size over all tasks. Similar techniques (e.g., random Fourier features, inducing points) have been used for scaling up MTGP training for large datasets. A special case where computation is reduced is when all tasks share the same sample size (i.e., $n_t = n$) and design points (i.e., $\mX_t = \mX$). Here, for certain kernels one can exploit the Kronecker structure of the Gram matrix for matrix inversion with $\calO(n^3 + T^3)$ work \cite{bonilla2007multi}. However, \cite{bonilla2007multi} also showed that in such case where simulations are assumed to be noise free, there is no ``information transfer'' between tasks, which defeats the purpose of multitask learning. There is thus a need for scalable MTGP modeling that can accommodate \textit{different design points} and \textit{sample sizes}, to enhance the benefits of multitask modeling. 

One limitation with the above scalable GP approaches is that they rely on some form of approximation of the predictive equations, where the quality of this approximation can be difficult to control in practice. For applications where the training inputs can be designed, there is a body of literature on structured GPs which explores pairings of design points with kernels for \textit{exact} and efficient computation of the predictive distribution. This includes the work of \cite{gardner2018product,wilson2015kernel}, which investigates the use of regular grid designs with product kernels; such grid designs, however, scale poorly with dimension. To address this issue, \cite{plumlee2014fast,wilson2014fast} explores the use of sparse grids and other partial grid structures; it is not clear, however, how such techniques extend for the multitask setting, where different designs and sample sizes are needed on each task. In the quasi-Monte Carlo (QMC) literature, there is a growing body of work on pairing low-discrepancy (LD) design points with specific kernels to induce desirable structure in the Gram matrix. For example, the use of a digitally-shifted digital sequence (e.g., the Sobol' sequence \cite{SOBOL196786}) with a digitally-shift-invariant kernel induces structure in the Gram matrix that permits efficient matrix inversion with $\calO(n \log n)$ work and $\calO(n)$ storage using the fast Walsh--Hadamard transform \cite{fino.fwht}. These pairings have been exploited for efficient single-task GP training in Bayesian cubature \cite{rathinavel.bayesian_QMC_sobol,rathinavel.bayesian_QMC_lattice,sorokin.FastBayesianMLQMC} and in solving random PDEs \cite{kaarnioja.kernel_interpolants_lattice_rkhs,sorokin.gp4darcy}.

\begin{wraptable}[16]{l}{0.55\textwidth}
    \centering
    \resizebox{0.55\textwidth}{!}{%
    \begin{tabular}{lcccc}
        \toprule
        & \multicolumn{2}{c}{computations} & \multicolumn{2}{c}{storage} \\
        \cmidrule(lr){2-3} \cmidrule(lr){4-5}
        & GP & fast GP & GP & fast GP \\
        \midrule
        single-task        & $n^3$ & $n \log n$                        & $n^2$ & $n$ \\
        multitask equal   & $N^3$ & $N \log N$                        & $N^2$ & $N$ \\
        multitask unequal & $N^3$ & $N \log N + N^2/n_\mathrm{min}$   & $N^2$ & $N^2/n_\mathrm{min}$ \\
        \bottomrule
    \end{tabular}
    }
    \caption{Comparison of cost and storage requirements for fitting GPs assuming the dimension $d=\calO(1)$ and the number of tasks $T=\calO(1)$ relative. The ``multitask unequal'' setting allows for arbitrary $n_1,\dots,n_T \in \bbN$ with $n_\mathrm{min} = \min\{n_1,\dots,n_T\}$, see \Cref{thm:costs_fast_mtgp} for the full cost. The ``multitask equal'' setting assumes $n_1=n_2=\dots=n_T := n$. ``Single-task'' has $T=1$ and $n_1 := n$, see \Cref{corr:costs_fast_mtgp_equal}. The fast GP constructions require that $n_t = 2^{m_t}$ for some $m_t \in \bbN_0$ for all $t \in \{1,\dots,T\}$.}
    \label{table:costs_storage}
\end{wraptable}

In this work, we propose a novel ``fast MTGP'' approach for efficient MTGP modeling with different design points and sample sizes for each task, which extends techniques on pairing LD designs with specific kernels to the multitask setting. Here, the required structured LD design points are desirable for two reasons. First, such designs are ``space-filling'' over the domain, which is beneficial for GP modeling \cite{gramacy2020surrogates,rasmussen.gp4ml}. Second, such designs have a random ``shifting'' mechanism that allows for the randomization of LD points; this has been broadly used in the field of randomized QMC \cite{owen.mc_book_practical,dick.digital_nets_sequences_book,dick2022lattice}. We will use these random shifts to generate \textit{different} designs (with potentially different sample sizes) within each task, and show that the corresponding Gram matrix admits desirable block structure when paired with specific multitask kernels. We then develop a new algorithm that exploits such structure for exact evaluation of the Gram matrix inverse and determinant, leveraging efficient techniques with fast Fourier or Walsh--Hadamard transforms along with careful recursive block matrix manipulations. We provide a theoretical analysis of the computational and storage costs for our algorithm, see the summary and comparison in \Cref{table:costs_storage}. 
Finally, we show the effectiveness of our fast MTGP approach on a suite of simulation experiments including our motivating QGP surrogate modeling application.


\section{Background \& Preliminaries} \label{sec:methods}

\subsection{Gaussian Processes} \label{sec:GPs}

Let $f: \calX \to \bbR$ denote the black-box function we wish to predict, e.g., the output of an expensive computer experiment. The fast GP and multitask GPs we consider will require $\calX = [0,1]^d$, the unit hypercube. Suppose $f$ follows a GP model, with constant mean $\tau \in \bbR$ and a symmetric positive definite covariance kernel $K: \calX \times \calX \to \bbR$. We will denote this by $f \sim \mathrm{GP}(\tau,K(\cdot,\cdot))$.

A popular choice for kernel $K$ is the squared exponential (or Gaussian) kernel $K(\bx,\bx') =  \gamma \exp\left(-\sum_{j=1}^d (x_j-x_j')^2/(2 \eta_j^2)\right)$ where $\gamma \in \bbR_+ := \{x \in \bbR: x >0\}$ is a variance parameter and $\bEta = (\eta_1, \dots, \eta_d) \in \bbR_+^d$ are per-dimension lengthscales. 

Suppose we have $n$ observations of the form $\by = \boldsymbol{f}+\bvarepsilon$, where $\boldsymbol{f} := (f(\bx_i))_{i=0}^{n-1}$ is the vector of true function values at design points $\mX := (\bx_i)_{i=0}^{n-1} \subseteq \calX$, and $\bvarepsilon \sim \calN(\bzero,\xi \mI_n)$ consists of IID zero-mean Gaussian noise terms. Here, the noise-free setting for deterministic computer experiments corresponds to the case of $\xi=0$. Conditional on such observations, one can show \cite{gramacy2020surrogates,rasmussen.gp4ml} that the posterior distribution of $f$ is also a Gaussian process, with the following closed-form expressions for its posterior mean and covariance
\begin{equation} \label{eq:GP_post_mean_cov} \begin{aligned}
    \bbE[f(\bx) | \mX, \by] &= \tau+\bK^\intercal(\bx) \tmK^{-1}(\by-\tau \bone), \qquad\text{and} \\ \Cov[f(\bx),f(\bx') | \mX] &= K(\bx,\bx') - \bK^\intercal(\bx) \tmK^{-1} \bK(\bx')
\end{aligned}
\end{equation}
respectively. 
Here $\bK(\bx) := (K(\bx,\bx_i))_{i=0}^{n-1} \in \bbR^n$ is a vector of kernel evaluations, and $\tmK := \mK + \xi \mI_n \in \bbR^{n \times n}$ where $\mK := (K(\bx_i,\bx_{i'}))_{i,i'=0}^{n-1} \in \bbR^{n \times n}$ is the Gram matrix.

For effective modeling, the model parameters of a GP (i.e., $\btheta = \{\tau,\gamma,\bEta,\xi\}$) need to be carefully estimated from data. Two popular approaches are maximum likelihood estimation (MLE; \cite{gramacy2020surrogates,rasmussen.gp4ml}) and generalized cross validation (GCV; \cite{wahba1990spline}), where $\btheta$ is selected to minimize the loss functions
\begin{equation} \label{eq:GP_MLE_GCV_loss}
    \calL_\mathrm{MLE}(\btheta) := (\by-\tau \bone)^\intercal \tmK^{-1} (\by-\tau \bone) + \log \lvert \tmK \rvert \qtq{and} \calL_\mathrm{GCV}(\btheta) := \tfrac{(\by - \tau \bone)^\intercal \tmK^{-2} (\by - \tau \bone)}{\left(\mathrm{trace}(\tmK^{-1})\right)^2},
\end{equation}
respectively. Here, the negative log-likelihood loss for MLE and the GCV loss are both expressed up to additive constants independent of the parameters $\btheta$. Note that the kernel matrix $\tmK$ implicitly depends on $\btheta$; we suppress this dependence for brevity. One can easily show \cite{gramacy2020surrogates,wahba1990spline,rasmussen.gp4ml} that the corresponding MLE and GCV estimators for the GP mean $\tau$ take the closed-form expressions
 \begin{equation} \label{eq:GP_optimal_tau}
        \htau_\mathrm{MLE} = \tfrac{\bone^\intercal \tmK^{-1} \by}{\bone^\intercal \tmK^{-1} \bone} \qtq{and}
        \htau_\mathrm{GCV} = \tfrac{\bone^\intercal \tmK^{-2} \by}{\bone^\intercal \tmK^{-2} \bone}.
    \end{equation}
Remaining parameters in $\btheta$ are estimated by minimizing the loss $\calL$ via numerical methods.  

The loss functions in \eqref{eq:GP_MLE_GCV_loss} show the key computational bottlenecks for GP model training: each evaluation of the loss function requires computing the matrix inverse $\tmK^{-1}$ and the matrix determinant $\lvert \tmK \rvert$ (for the MLE loss). 
Standard GP fitting requires $\calO(n^3+n^2d)$ work and $\calO(n^2)$ storage, under our standing assumption that  evaluation of the kernel $K$ requires $\calO(d)$ work. \Cref{table:costs_storage} summarizes these requirements assuming $d=\calO(1)$. These costs may be prohibitive when the sample size $n$ grows large. 

\subsection{Fast Gaussian Processes} \label{sec:FGPs}

As reviewed in the introduction, there is a growing literature on scaling up GP training for large sample sizes $n$, with many methods employing different approximations of the predictive equations in \eqref{eq:GP_post_mean_cov}. To achieve an \textit{exact} evaluation of these equations, one strategy is to adopt a careful pairing of ``low-discrepancy'' design points $\mX$ and kernel $K$, to induce desirable structure in the kernel matrix $\tmK$ for efficient matrix inversion and determinant calculation. These ``fast GP'' approaches were originally developed in \cite{zeng.spline_lattice_digital_net,zeng.spline_lattice_error_analysis}, and recently applied to fast Bayesian cubature \cite{rathinavel.bayesian_QMC_lattice,rathinavel.bayesian_QMC_sobol,sorokin.FastBayesianMLQMC} and fast PDE solvers \cite{kaarnioja.kernel_interpolants_lattice_rkhs,sorokin.gp4darcy}. We briefly review the low-discrepancy design and kernel choices for fast GPs in the following.

\subsubsection{Low-discrepancy designs: lattices and digital sequences} \label{sec:LD_points_intro}

Low-discrepancy design points aim to reduce the discrepancy between the empirical distribution of design points and the uniform distribution over $\calX = [0,1]^d$. Such designs were originally developed in the quasi-Monte Carlo literature for integration (with provably faster convergence over Monte Carlo methods), but have been widely adopted for other uses, e.g., as experimental designs for computer experiments \cite{fang2005design,sakai2026space}.

\begin{figure}[!t]
    \centering 
    \includegraphics[width=1\textwidth]{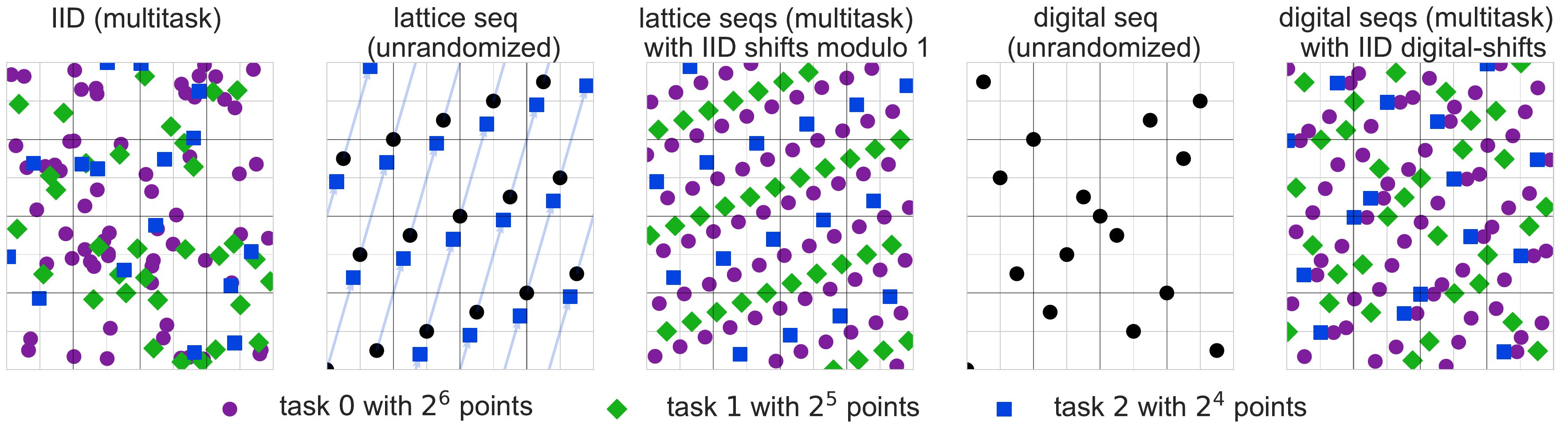}
    \caption{IID points, LD shifted rank-1 lattice sequences, and LD base-2 digitally-shifted digital sequences. In the multitask plots, each sequence has three distinct shifts in different colors, one for each task. The LD lattice has three distinct shifts (modulo one), while the LD digital sequence has three distinct digital-shifts. Different numbers of points are used for different tasks. Notice the more uniform coverage of LD sequences compared to IID points.}
    \label{fig:points_full}
    \vspace{-.75cm}
\end{figure}

The two LD designs primarily considered in the fast GP literature are shifted rank-$1$ lattices \cite{dick2022lattice} and digitally-shifted digital sequences \cite{dick.digital_nets_sequences_book} (of which the Sobol' \cite{SOBOL196786} sequence is a special case), with both cited books providing thorough theoretical treatments. We refer to \cite{choi.QMC_software,hickernell.qmc_what_why_how,owen.mc_book_practical,sorokin.2025.ld_randomizations_ho_nets_fast_kernel_mats} for accessible introductions to both LD designs along with intuition on their constructions, randomizations, and uniformity properties. \Cref{fig:points_full} shows randomizations of both types of LD points contrasted with IID design points sampled uniformly over $\calX$. Notice that both LD designs visually ``fill'' the domain space in a more even manner than IID points, which is unsurprising since these designs have provably lower discrepancies. Such a ``space-filling'' property not only enables faster convergence of quasi-Monte Carlo methods for integration, but is also desirable for the design of computer experiments \cite{gramacy2020surrogates}. Moreover, unlike regular grids which require an exponential sample size in the dimension, both LD sequences considered here attain high uniformity properties at sample sizes that are powers of two.

The first class of LD points for fast GPs are shifted rank-$1$ lattices \cite{dick2022lattice} which we formally define below; we consider these only in base two. 

\begin{definition}[Shifted rank-$1$ lattice] \label{def:lattices}
    For a fixed generating vector $\bg \in \bbN^d$, the rank-$1$ lattice $(\bz_i)_{i \geq 0} \subset [0,1)^d$ in radical-inverse order has $\bz_i = v(i) \bg \bmod 1$ where $v(i)$ is the van der Corput sequence~\cite{vandercorput} with $v(0)=0$ and $v(i) = \sum_{p=0}^{\lfloor \log_2(i) \rfloor} \mi_p 2^{-p-1}$ for $i=\sum_{p=0}^{\lfloor \log_2(i) \rfloor } \mi_p 2^p$. For a shift $\bDelta \in [0,1)^d$, the shifted rank-$1$ lattice $(\bx_i)_{i \geq 0}$ has $\bx_i = (\bz_i + \bDelta) \bmod 1$.
\end{definition}

In \Cref{fig:points_full} we see the shift modulo one mechanism for lattices.

The second class of LD points for fast GPs are digitally-shifted digital sequences \cite{dick.digital_nets_sequences_book}. Here we will only consider base two digital sequences for simplicity. The definition of these LD points relies on the digital addition operator $\oplus$ which performs digit-wise addition modulo the base. Specifically, for scalars $a,b \in [0,1)$ with terminating binary expansions (called dyadic rationals) $a = \sum_{p=1}^\infty \ma_p 2^{-p}$ and $b = \sum_{p=1}^\infty \mb_p 2^{-p}$, we set $a \oplus b = \sum_{p=1}^\infty ((\ma_p + \mb_p) \bmod 2) 2^{-p}$ to be the exclusive or (XOR) between binary digits. For dyadic rational vectors $\ba,\bb \in [0,1)^d$, we set $\ba \oplus \bb := (a_1 \oplus b_1,\dots,a_d \oplus b_d)$ to act elementwise. Digitally-shifted digital sequences are formally defined as follows.

\begin{definition}[Base-2 digitally-shifted digital sequences] \label{def:dseqs}
    For generating matrices $\mG \in [0,1)^{d \times \infty}$ of dyadic rationals with zeroth column $\bg_0$ and $p$-th column $\bg_p$, the digital sequence $(\bz_i)_{i \geq 0} \subset [0,1)^d$ in  radical-inverse order has $\bz_i = \bigoplus_{p=0}^{\lfloor\log_2(i)\rfloor} \mi_p \bg_p$ where $i=\sum_{p=0}^{\lfloor \log_2(i) \rfloor } \mi_p 2^p$. For a digital-shift $\bDelta \in [0,1)^d$, the digitally-shifted digital sequence $(\bx_i)_{i \geq 0}$ has $\bx_i = \bz_i \oplus \bDelta$.
\end{definition}

The digital-shifting mechanism optionally swaps neighboring intervals of size $2^{-p}$ for each dimension. For example, in $d=1$ dimension, the first bit in $\Delta$ determines if $[0,1/2)$ swaps with $[1/2,1)$, the second bit determines if $[0,1/4)$ swaps with $[1/4,1/2)$ and $[1/2,3/4)$ swaps with $[3/4,1)$, and so on. Digital sequences balance products of such intervals with equal numbers of points. For example, in \Cref{fig:points_full} the unrandomized digital sequence has one point in each $1/4 \times 1/4$ tile. Such balance properties are retained under digital-shifts. 

Of course the dyadic rational requirement is not an issue for our finite precision implementation. Moreover, despite digital sequences living in $[0,1)^d$ and the digital-shift operator $\oplus$ acting on points in $[0,1)^d$, we practically accommodate the entire $[0,1]^d$ domain by mapping points equal to $1$ to the the largest value less than $1$ in the desired floating-point precision. In what follows we will often list the domain of digital operations as $[0,1]^d$ with this boundary mapping understood. A similar comment holds for lattices, where points equal to $1$ are implicitly identified as being equal to $0$ via the modulo $1$ wrapping. 

\subsubsection{Shift-invariant and digitally-shift-invariant kernels} \label{sec:FGP_SI_DSI_kernels} 

Next, we review certain kernel choices that pair well with the introduced LD design points for efficient matrix inversion. These kernels will adopt the following product form
\begin{equation} \label{eq:prod_kernel} 
    K(\bx,\bx') = \gamma {\textstyle \prod_{j=1}^d} \left[1+\eta_j \calK(x_j,x_j')\right]
\end{equation}
where $\calK: [0,1] \times [0,1] \to \bbR$ is a univariate kernel. In this form, the parameters $\bEta = (\eta_1, \dots, \eta_d)$ model variable importance, where a larger $\eta_j > 0$ (estimated from data) reflects greater importance for variable $j$.
Such forms are full ANOVA kernels with product (first-order) weights \cite{wahba1990spline,duvenaud2011additive}, which are commonly used in the the quasi-Monte Carlo and fast GP literature \cite{kaarnioja.kernel_interpolants_lattice_rkhs,kuo.application_qmc_elliptic_pde,rathinavel.bayesian_QMC_lattice,rathinavel.bayesian_QMC_sobol,sorokin.FastBayesianMLQMC,sorokin.gp4darcy}. Higher-order ANOVA kernels may be accommodated in the fast GP framework, although then kernel evaluation is not necessarily $\calO(d)$ as we assume here.  

The first case of a shifted lattice design $(\bx_i)_{i \in \bbN_0}$ (see \Cref{def:lattices}) is typically paired with a shift-invariant (SI) kernel in the fast GP literature. This pairing (which we call ``SI-LAT'') induces a permuted circulant structure for the Gram matrix $\tmK$, where its eigenvectors are a bit-reversal permutation matrix times the inverse discrete Fourier matrix (more on this later). Such structure may be exploited for efficient matrix inversion and determinant computation with $\calO(n \log n)$ cost using fast Fourier transforms (FFTs; \cite{cooley1965algorithm}). A shift-invariant kernel is formally defined as

\begin{definition}[Shift-invariant (SI) kernel] \label{def:si_kernel}
    A kernel $K: [0,1]^d \times [0,1]^d \to \bbR$ is SI when $K(\bx,\bx') = \hK((\bx-\bx')\bmod 1)$ for some $\hK: [0,1]^d \to \bbR$ and any $\bx,\bx' \in [0,1]^d$. 
\end{definition}

One popular class of SI kernels takes the product form in \eqref{eq:prod_kernel} with univariate kernels 
\begin{equation} \label{eq:SI_kernel_1d_ex}
    \calK(x,x') = \tfrac{(-1)^{\alpha+1}(2 \pi)^{2 \alpha}}{(2\alpha)!} B_{2\alpha}((x-x') \bmod 1),
\end{equation}
where $\alpha \in \bbN$ is a smoothness parameter and $B_p$ denotes the $p$-th Bernoulli polynomial. Such kernels have been extensively studied throughout the QMC literature \cite{kuo.application_qmc_elliptic_pde,kaarnioja.kernel_interpolants_lattice_rkhs,dick2022lattice}. 

A potential drawback of SI kernels is that they are necessarily periodic. One can show that the kernel form \eqref{eq:SI_kernel_1d_ex} has $\alpha-1$ periodic derivatives, e.g., the kernel with $\alpha=1$ is periodic, whereas the kernel with $\alpha=2$ is periodic with a periodic derivative. This restriction may limit the applicability of such kernels in practical applications where the black-box function is not necessarily periodic, or where periodizing transforms either destroy smoothness or introduce high model variation. The following digitally-shift-invariant (DSI) kernels do not have this periodicity restriction, but do include discontinuities. 

The second case of a digitally-shifted base-$2$ digital sequences $(\bx_i)_{i \in \bbN_0}$ (see \Cref{def:dseqs}) is typically paired with a digitally-shift-invariant kernel $K$ in the fast GP literature. This pairing (which we call ``DSI-DSEQ'') induces a Hadamard structure for the Gram matrix $\mK$, where its eigenvectors are the normalized Walsh--Hadamard matrix. Such structure can similarly be exploited for efficient matrix inversion and determinant computation with $\calO(n \log n)$ cost using fast Walsh--Hadamard transforms (FWHTs; \cite{fino.fwht}). A DSI kernel is defined as

\begin{definition}[Digitally-shift-invariant (DSI) kernel] \label{def:dsi_kernel}
    A kernel $K: [0,1]^d \times [0,1]^d \to \bbR$ is DSI when $K(\bx,\bx') = \hK(\bx \oplus \bx')$ for some $\hK: [0,1]^d \to \bbR$ and any $\bx,\bx' \in [0,1]^d$.  
\end{definition}

One form of a DSI kernel is the adaptive-smoothness DSI product kernel \cite{sorokin.FastBayesianMLQMC}, which takes the product form in \eqref{eq:prod_kernel} with
\begin{equation} \label{eq:DSI_kernel_1d_ex}
    \begin{aligned} 
    \calK(x,x') &= b_1 \hcalK_1(x \oplus x') + b_2 \hcalK_2(x \oplus x') + b_3 \hcalK_3(x \oplus x') + b_4 \hcalK_4(x \oplus x') \qquad\text{and } \\
    \hcalK_\alpha(x) &= \begin{cases} \tfrac{1}{6}\left(1-\tfrac{1}{2}t_1(x)\right), & \alpha=1 \\ -1 -\beta(x) x + \tfrac{5}{2}\left[1-t_1(x)\right], & \alpha = 2 \\ -1+\beta(x)x^2-5\left[1-t_1(x)\right]x+\tfrac{43}{18}\left[1-t_2(x)\right], & \alpha = 3 \\ -1 -\tfrac{2}{3}\beta(x)x^3+5\left[1-t_1(x)\right]x^2 - \tfrac{43}{9}\left[1-t_2(x)\right]x \\
    \quad +\tfrac{701}{294}\left[1-t_3(x)\right]+\beta(x)\left[\tfrac{1}{48} \sum_{a=1}^\infty \tfrac{1-2\mx_a}{2^{3(a-1)}} - \tfrac{1}{42}\right], & \alpha = 4, \end{cases}
    \end{aligned}.
\end{equation}
Here $x$ has binary expansion $x = \sum_{p=1}^\infty \mx_p 2^{-p}$ where $\mx_p \in \{0,1\}$ are the binary digits, $\beta(x) = - \lfloor \log_2(x) \rfloor$ and $t_\nu(x) = 2^{-\nu \beta(x)}$ with the convention $\beta(0) = t_\nu(0) = 0$, and $\bb = (b_1, \cdots, b_4) \in \bbR_+^4$ are hyperparameter weights to be optimized. Again, $\alpha$ is a smoothness parameter, with the $\alpha=1$ kernel $\hcalK_1$ corresponding to the order-one Walsh space in \cite{dick.multivariate_integraion_sobolev_spaces_digital_nets}, and the higher-order $\alpha>1$ kernels $\hcalK_\alpha$ from \cite{baldeaux.HO_nets_RKHS} ($\alpha \in \{2,3\}$)  and \cite{sorokin.2025.ld_randomizations_ho_nets_fast_kernel_mats} ($\alpha=4$). It has been shown that certain Sobolev spaces are continuously embedded into the reproducing kernel Hilbert spaces corresponding to certain higher-order $\alpha>1$ DSI kernels $\hcalK_\alpha$. Recent numerical experiments in \cite{sorokin.fastgps_probnum25} showed the adaptive smoothness DSI kernel in \eqref{eq:DSI_kernel_1d_ex} gives competitive performance compared to Gaussian kernels across a range of financial modeling and PDE benchmark problems.

\subsubsection{Fast GP computations} 

Using these design-kernel pairings, one can exploit the induced structure on the Gram matrix for efficient matrix calculations. The following condition summarizes these pairing requirements:

\begin{condition}[Fast GP] \label{cond:FGP}
    Suppose the sample size is $n=2^m$ for some $m \in \bbN_0$, and either  
    \begin{enumerate}
        \item (\textbf{SI-LAT}) $K$ is SI (\Cref{def:si_kernel}) and $\mX$ is a shifted rank-$1$ lattice in radical-inverse order (\Cref{def:lattices}), or 
        \item (\textbf{DSI-DSEQ}) $K$ is DSI (\Cref{def:dsi_kernel}) and $\mX$ is a digitally-shifted base-$2$ digital sequence in radical inverse order (\Cref{def:dseqs}).
    \end{enumerate}
\end{condition}

Under the first SI-LAT pairing, the Gram matrix takes a (permuted) circulant form which allows for efficient matrix inversion and  determinant computations via FFTs with $\calO(n \log n)$ work. Similarly, under the DSI-DSEQ pairing, the Gram matrix takes a structured form which allows for efficient matrix inversion and determinant computations via fast Walsh--Hadamard transforms (FWHT) also with $\calO(n \log n)$ work. More specifically, in either case the eigendecomposition of the Gram matrix has an eigenvector matrix corresponding to the FFT / FWHT. Eigenvalues may be computed by taking the fast transform applied to the first column of the Gram matrix, hence enabling fast GP's linear storage requirement in $n$.  Moreover, the estimated means are exactly the sample means. These results are formalized in the following theorem from \cite{rathinavel.bayesian_QMC_sobol,rathinavel.bayesian_QMC_lattice}. 

\begin{theorem}[Fast GP computations] \label{thm:FGP_computations}
    Under \Cref{cond:FGP}, we have the eigendecomposition $\tmK = \mV \tmLambda \, \mV^\dagger$ where:
    \begin{enumerate}
        \item $\mV$ is unitary and has the constant zeroth column $\mV_{:,0} = \bone/\sqrt{2^m}$ and zeroth row $\mV_{0,:} = \bone^\intercal/\sqrt{2^m}$. 
        \item Computing $\mV \ba$, $\mV^\dagger \ba$, $\tmK \ba$, $\tmK^{-1} \ba$, and $\lvert \tmK \rvert$ for any $\ba \in \bbC^n$ each cost $\calO(2^m m)$. 
        \item $\tblambda := \tmLambda \bone = \sqrt{n} \, \mV^\dagger \, \tmK_{:,0}$ where $\tmK_{:,0}$ is the zeroth column of $\tmK$. 
        \item $\htau_\mathrm{MLE} = \htau_{GCV} = \bone^\intercal \by / n$ in \eqref{eq:GP_optimal_tau}, i.e., the optimal constant means are the sample means. 
    \end{enumerate}
    For the SI-LAT condition, $\mV^\dagger \ba$ applies a bit-reversal permutation to $\ba$ then applies an FFT to the result. For the DSI-DSEQ condition, $\mV^\dagger \ba$ applies a FWHT to $\ba$ and $\mV$ is real so $\mV^\dagger = \mV$. 
\end{theorem}

While this body of work on fast GP approaches is promising, there has been little work on extending these techniques for the multitask setting, where similar issues of scalability are present. The extension of such techniques require new modeling and computational algorithms to tackle the need for different design points and different sample sizes within different tasks. We propose a novel fast MTGP approach to tackle these limitations.

\section{Proposed Fast Multitask Gaussian Processes} \label{sec:MTGPs}

Let us first describe the multitask GP framework before outlining the proposed pairing of design points and kernels in the fast multitask setting. With such pairings, we then present the proposed algorithm for efficient matrix calculations, and prove its computational and storage complexities.

\subsection{Multitask Setting} \label{sec:MTGP_setup}

We consider the following multitask setting. Suppose we wish to predict $T > 1$ different but related black-box functions (i.e., tasks), each with common domain $\calX$. 
We denote these functions by $f: \{1,\dots,T\} \times \calX \to \bbR$, where $f(t,\bx)$ is the output of the $t$-th task function at inputs $\bx \in \calX$. We wish to exploit dependencies between tasks for effective multitask predictions.

One way to do this is using a multitask GP model. Suppose $f$ follows a multitask GP model, i.e., $f \sim \mathrm{GP}(\tau(\cdot),K(\cdot,\cdot))$. Here $K: (\{1,\dots,T\} \times \calX) \times (\{1,\dots,T\} \times \calX) \to \bbR$ is a symmetric positive definite covariance kernel over the joint space of tasks and inputs. We take the mean function $\tau$ to be different constants for each task, and thus identify it with $\btau = (\tau_1,\dots,\tau_T)$ such that $\bbE[f(t,\bx)] = \tau_t$ for all $\bx \in \calX$ and any $t \in \{1,\dots,T\}$.

In what follows, we adopt a product form for the multitask GP kernel $K$, which is widely used in the multitask GP literature \cite{bonilla2007multi}. Specifically, we presume $K$ takes the form 
\begin{equation} \label{eq:mtprod}
    K((t,\bx),(t',\bx')) = R(t,t') Q(\bx,\bx'),
\end{equation}
where $R: \{1,\dots,T\} \times \{1,\dots,T\} \to \bbR$ is a kernel that models task dependencies, and $Q$ is a kernel that models input dependencies such as a standard Gaussian kernel or the product form kernel in \eqref{eq:prod_kernel}. A common strategy for modeling the task kernel $R$ is to directly parameterize $\mR := (R(t,t'))_{t,t'=1}^T = \mB \mB^\intercal + \mathrm{diag}(\bzeta)$, where $\mB \in \bbR^{T \times s}$ with $s \leq T$ is a low-rank matrix and $\bzeta \in \bbR_+^T$ is a diagonal vector. In our numerical experiments, we will use full rank $s=T$ and treat $\mB$ and $\bzeta$ as hyperparameters to be optimized with MLE or GCV. 

Suppose, for the $t$-th task, we observe data of the form $\by_t = \boldsymbol{f}_t + \bvarepsilon_t$. As before, $\boldsymbol{f}_t = (f(t,\bx_{t i}))_{i=0}^{n_t-1}$ is the vector of true function values at design points $\mX_t := (\bx_{t i})_{i=0}^{n_t-1} \subseteq \calX$, where $\bvarepsilon_t \sim \calN(\bzero,\xi_t \mI_{n_t})$ consists of IID zero-mean Gaussian noise terms. Here, $n_t$ samples are obtained for the $t$-th task, and $N = \sum_{t=1}^T n_t$ samples are observed over all tasks. The noise vectors $\bvarepsilon_1,\dots,\bvarepsilon_T$ are also assumed to be independent.

Conditional on such data, one can show that the posterior distribution of $f$ given design points $\mX := (\mX_t)_{t=1}^T \in \calX^N$ and noisy observations $\by = (\by_t)_{t=1}^T \in \bbR^N$ is also a Gaussian process. The following closed-form expressions for its posterior mean and covariance
\begin{equation} \label{eq:MTGP_post_mean_cov}
    \begin{aligned}
        \bbE[f(t,\bx) | \mX, \by] &= \tau_t+\bK^\intercal(t,\bx) \tmK^{-1}(\by-\mE^\intercal \btau) \qquad\text{and} \\
        \Cov[f(t,\bx),f(t',\bx') | \mX] &= K((t,\bx),(t',\bx')) - \bK^\intercal(t,\bx) \tmK^{-1} \bK(t',\bx')
    \end{aligned}
\end{equation}
respectively follow from their single-task counterparts in \eqref{eq:GP_post_mean_cov}. Here, \patchoverfullhbox $\bK(t,\bx) := (\bK_{t'}(t,\bx))_{t'=1}^T \in \bbR^N$, $\bK_{t'}(t,\bx) := (K((t,\bx),(t',\bx_{t' i})))_{i=0}^{n_{t'}-1} \in \bbR^{n_{t'}}$, and $\tmK := (\tmK_{t t'})_{t,t'=1}^T \in \bbR^{N \times N}$ where $\tmK_{t t'} := \mK_{t t'} + \delta_{t t'} \xi_t \mI_{n_t}$, $\mK_{t t'} := (K((t,\bx_{t i}),(t',\bx_{t' i'})))_{i,i'=0}^{n_t-1,n_{t'}-1}$, and $\delta_{t t'}$ is an indicator function equaling $1$ if $t=t'$ and $0$ otherwise. In \eqref{eq:MTGP_post_mean_cov}, $\mE \in \bbR^{T \times N}$ is the ``task-summing'' matrix given by $\mE = \begin{bsmallmatrix} \bone_1^\intercal & & \\ & \ddots & \\ & & \bone_T^\intercal \end{bsmallmatrix}$, for example when $T=2$ with $n_1=2$ and $n_2=3$ we have $\mE = \begin{bsmallmatrix} 1 & 1 & 0 & 0 & 0 \\ 0 & 0 & 1 & 1 & 1 \end{bsmallmatrix} \in \bbR^{2 \times 5}$.

The model parameters of this multitask GP (i.e., $\btheta = \{\btau,\gamma,\bEta,\bxi,\mB,\bzeta\}$) similarly needs to be carefully estimated from data. One can show that the MLE and GCV loss functions for the MTGP take the form
\begin{equation}
    \calL_\mathrm{MLE}(\btheta) := (\by-\mE^\intercal \btau)^\intercal \tmK^{-1} (\by-\mE^\intercal \btau) + \log \lvert \tmK \rvert \quad\text{and}\quad \calL_\mathrm{GCV}(\btheta) := \tfrac{(\by - \mE^\intercal \btau)^\intercal \tmK^{-2} (\by - \mE^\intercal \btau)}{\mathrm{trace}^2(\tmK^{-1})}. \label{eq:MTGP_MLE_GCV_loss} 
\end{equation}
following their single-task counterparts in \eqref{eq:GP_MLE_GCV_loss}. These are respectively minimized by the mean estimates $\hbtau_\mathrm{MLE}$ and $\hbtau_\mathrm{GCV}$ which satisfy
\begin{equation} \label{eq:MTGP_optimal_tau}
    (\mE \tmK^{-1} \mE^\intercal) \hbtau_\mathrm{MLE} = \mE \tmK^{-1} \by \qtq{and}
    (\mE \tmK^{-2} \mE^\intercal) \hbtau_\mathrm{GCV} = \mE \tmK^{-2} \by.
\end{equation}

As before, the key computational bottleneck for MTGP training lies in evaluating the above loss function, which involves evaluating the matrix inverse $\tmK^{-1}$ and the matrix determinant $\lvert \tmK \rvert$. The required computational and storage costs in this multitask setting are $\calO(N^3 + N^2d)$ and $\calO(N^2)$, respectively, which is analogous to the earlier single-task setting. \Cref{table:costs_storage} summarizes these costs, which can again be prohibitively expensive for MTGP training when either $T$ (the number of tasks) grows large or $n_t$ (the sample size within a task) grows large. 

\subsection{Multitask Design Points and Kernels} \label{sec:FMTGP_framework}

For efficient multitask GP modeling, we employ specific pairings of multitask design points and kernels, extending the literature on fast GPs. For each task, we make use of a low-discrepancy design points from \Cref{sec:LD_points_intro}. Here, we allow for varying sample sizes over different tasks, e.g., task 1 can have $2^8$ design points, whereas task 2 can have $2^7$ design points. We further allow for varying design points within each task, by exploiting the random \textit{shifting} mechanism for the introduced low-discrepancy points. In particular, even if all tasks share the same sample size, the design points for each task may use a \textit{different} random shift vector $\bDelta_t$, which guarantees different design points for different tasks, see \Cref{fig:points_full}. 


For kernel choice, we make use of the product kernel form in \eqref{eq:mtprod}, where the input kernel $Q$ is selected based on the choice of design points. If the multitask design points are generated from a shifted lattice rule, then $Q$ is selected as a shift-invariant kernel (see \Cref{def:si_kernel}). If the multitask design points are generated instead from a digitally-shifted digital sequence, then $Q$ is selected as a digitally-shift-invariant kernel (see \Cref{def:dsi_kernel}). With such a pairing of multitask design points and kernel, we can induce desirable structure in the multitask Gram matrix $\tmK$ that can be exploited for efficient matrix computations.

We summarize these two multitask design-kernel pairing conditions for our fast MTGP approach in the following.

\begin{condition}[Fast MTGP] \label{cond:FMTGP}
    Suppose $\calX = [0,1]^d$ and \patchoverfullhbox $K: (\{1,\dots,T\} \times [0,1]^d) \times (\{1,\dots,T\} \times [0,1]^d) \to \bbR$ is a multitask product kernel with task kernel $R: \{1,\dots,T\} \times \{1,\dots,T\} \to \bbR$ and input kernel $Q: [0,1]^d \times [0,1]^d \to \bbR$. Suppose $n_t=2^{m_t}$ for some $m_t \in \bbN_0$ for any $t \in \{1,\dots,T\}$, and assume $n_1 \geq \cdots \geq n_T$, which may always be satisfied by reindexing tasks. Suppose either
    \begin{enumerate}
        \item (\textbf{SI-LAT}) $Q$ is SI (\Cref{def:si_kernel}) and $\mX_1,\dots,\mX_T$ are shifted rank-$1$ lattices in radical-inverse order (\Cref{def:lattices}) with a common generating vector $\bg \in \bbN^d$ and possibly different shifts per task, i.e., $\bx_{t i} = (v(i) \bg + \bDelta_t) \bmod 1$ for some $\bDelta_1,\dots,\bDelta_T \in [0,1]^d$, or 
        \item (\textbf{DSI-DSEQ}) $Q$ is DSI (\Cref{def:dsi_kernel}) and $\mX_1,\dots,\mX_T$ are digitally-shifted base-$2$ digital sequences in radical-inverse order (\Cref{def:dseqs}) with common generating matrices $\mG \in [0,1)^{d \times \infty}$ and possibly different digital-shifts per task, i.e., $\bx_{t i} = \bigoplus_{p=0}^{\lfloor \log_2(i) \rfloor} \mi_p \bg_p \oplus \bDelta_t$ for some $\bDelta_1,\dots,\bDelta_T \in [0,1]^d$.
    \end{enumerate}
\end{condition} 

As discussed earlier in \Cref{sec:FGP_SI_DSI_kernels}, SI kernels (and thus the SI-LAT pairing) are appropriate when the black-box functions for each task are known to be periodic. When this is not the case, one should instead employ the DSI-DSEQ pairing, since the DSI kernels permit non-periodicity for function modeling. Because of this, our numerical experiments later in \Cref{sec:numerics} will only consider the DSI-DSEQ pairing.

\subsection{Algorithm for Gram Matrix Calculations} \label{sec:FMTGP_algorithm}

With the multitask design-kernel pairings in \Cref{cond:FMTGP}, we can exploit useful structure on $\tmK$ for efficient block Gram matrix calculations, i.e., its matrix inversion and determinant computation. In \Cref{thm:FMTGP_computations} below we show that using a product kernel with a SI/DSI input kernel paired to matching LD sequences give block Gram matrices with structured blocks of different sizes. After factoring out certain block diagonal matrices with FFT/FWHT blocks from both sides, it is sufficient to invert and compute the determinant of a highly sparse block diagonal matrix with blocks of different sizes. \Cref{algo:inv_det_Theta} accomplish this via block Gaussian elimination under a certain task ordering which yields diagonal Schur complements, thus enabling efficient computation. The computational complexity and storage required by \Cref{algo:inv_det_Theta} are analyzed in \Cref{thm:costs_fast_mtgp}, with \Cref{corr:costs_fast_mtgp_equal} giving the special case of equal sample sizes across tasks. \Cref{table:costs_storage} summarizes these costs in the setting where $d = \calO(1)$ and $T = \calO(1)$. We begin by generalizing \Cref{thm:FGP_computations} to the following theorem which develops the structure in each block of the Gram matrix $\tmK_{tt'} \in \bbR^{n_t \times n_{t'}}$ where $t,t' \in \{1,\dots,T\}$. The proof is a straightforward generalization of the computations in \cite{rathinavel.bayesian_QMC_lattice,rathinavel.bayesian_QMC_sobol} after noting that, for $2^p > i$, the shifted rank-1 lattice in \Cref{def:lattices} satisfies $\bx_{2^p+i} = (\bx_{2^p} + \bx_i) \bmod 1$ and the digitally-shifted digital sequences in \Cref{def:dseqs} satisfy $\bx_{2^p+i} = \bx_{2^p} \oplus \bx_i$. 

\begin{theorem}[Fast MTGP computations] \label{thm:FMTGP_computations}
    Under \Cref{cond:FMTGP}, we have the decomposition $\tmK_{t t'} = \mV_{m_t} \tmLambda_{t t'} \mV_{m_{t'}}^\dagger \in \bbR^{2^{m_t} \times 2^{m_{t'}}}$, where, for all $m \in \bbN_0$, 
    \begin{enumerate}
        \item $\mV_m \in \bbC^{2^m \times 2^m}$ is unitary and has the constant zeroth column $(\mV_m)_{:,0} = \bone/\sqrt{2^m}$ and zeroth row $(\mV_m)_{0,:} = \bone^\intercal/\sqrt{2^m}$.
        \item Computing $\mV_m \ba$ and $\mV_m^\dagger \ba$ for any $\ba \in \bbC^{2^m}$ each cost $\calO(2^m m)$. 
        \item Regarding $\tmLambda_{t t'} \in \bbC^{2^{m_t} \times 2^{m_{t'}}}$, 
        \begin{itemize}[leftmargin=*]
            \item (tall-case) if $m_t \geq m_{t'}$ then $\tmLambda_{t t'}$ is a $2^{m_t - m_{t'}} \times 1$ block matrix with $2^{m_{t'}} \times 2^{m_{t'}}$ diagonal blocks which is fully specified by $\tblambda_{t t'} := \tmLambda_{t t'} \bone = \sqrt{2^{m_{t'}}} \; \mV_{m_t}^\dagger (\tmK_{t t'})_{:,0}$ where $(\tmK_{t t'})_{:,0}$ is the zeroth column of $\tmK_{t t'}$, whereas 
            \item (wide-case) if $m_t \leq m_{t'}$ then $\tmLambda_{t t'}$ is a $1 \times 2^{m_{t'} - m_t}$ block matrix with $2^{m_t} \times 2^{m_t}$ diagonal blocks which is fully specified by $\tblambda_{t t'} := \tmLambda_{t t'}^\intercal \bone = \sqrt{2^{m_t}} \mV_{m_{t'}}^\intercal (\tmK_{t t'})_{0,:}^\intercal$ where $(\tmK_{t t'})_{0,:}$ is the zeroth row of $\tmK_{t t'}$. 
        \end{itemize}
        \item The MLE and GCV mean constant estimates in \eqref{eq:MTGP_optimal_tau} respectively satisfy $\mPi \hbtau_\mathrm{MLE} = \mE \tmK^{-1} \by$ and $(\mH\mH^\intercal) \hbtau_\mathrm{GCV} = \mE \tmK^{-2} \by$ where $\mPi := \left(\sqrt{n_t n_{t'}} \left(\tmLambda^{-1}\right)_{N_{t-1},N_{t'-1}}\right)_{t,t'=1}^T \in \bbR^{T \times T}$ and $\mH := \left(\sqrt{n_t} \left(\tmLambda^{-1}\right)_{N_{t-1},k n_T}\right)_{t=1,k=0}^{T,N/n_T-1} \in \bbC^{T \times N / n_T}$ where $N_0 := 0$ and $N_t := n_1 + \dots + n_t$. 
        \item If $\tmK$ is Hermitian positive definite (HPD), then $\tmLambda_{:t,:t}$ is HPD (and in particular $\tmLambda = \tmLambda_{:T,:T}$ is HPD), and $\tmLambda_{tt'} = \tmLambda_{t't}^\dagger$ for all $t,t' \in \{1,\dots,T\}$.
    \end{enumerate}
    Under the SI-LAT condition, $\mV_m^\dagger \ba$ applies a bit-reversal permutation to $\ba$ then applies an FFT to the result. Under the DSI-DSEQ condition, $\mV_m^\dagger \ba$ applies a FWHT to $\ba$ with real $\mV_m$ so $\mV_m^\dagger = \mV_m$. 
\end{theorem}

To compute the inverse and determinant of $\tmK$, it is sufficient to compute the inverse and determinant of the sparse matrix $\tmLambda := (\tmLambda_{t t'})_{t,t'=1}^T$ as $\tmK = \mV \tmLambda\, \mV^\dagger$ where $\mV := \diag(\mV_{m_1},\dots,\mV_{m_T})$ is a block diagonal matrix of fast transforms. We will use the notation $\tmLambda_{:t,:t'} := (\tmLambda_{\dell \dell'})_{\dell,\dell'=1}^{t,t'}$ to denote the first $t$ row blocks and first $t'$ column blocks, with $\tmLambda_{:T,:T} = \tmLambda$. We will also use $\tmLambda_{t,:t'} = (\tmLambda_{t \dell'})_{\dell'=1}^{t'}$ to denote the first $t'$ column blocks in the $t$-th row, and use $\tmLambda_{:t',t} = (\tmLambda_{\dell' t})_{\dell'=1}^{t'}$ to denote the first $t'$ row blocks in the $t$-th column. Let us also write  $\tmLambda_{:t,:t}^{-1} := (\tmLambda_{:t,:t})^{-1}$ so $\tmLambda^{-1} = \tmLambda_{:T,:T}^{-1}$. 

\begin{figure}[!t]
    \centering 
     \includegraphics[width=1\textwidth]{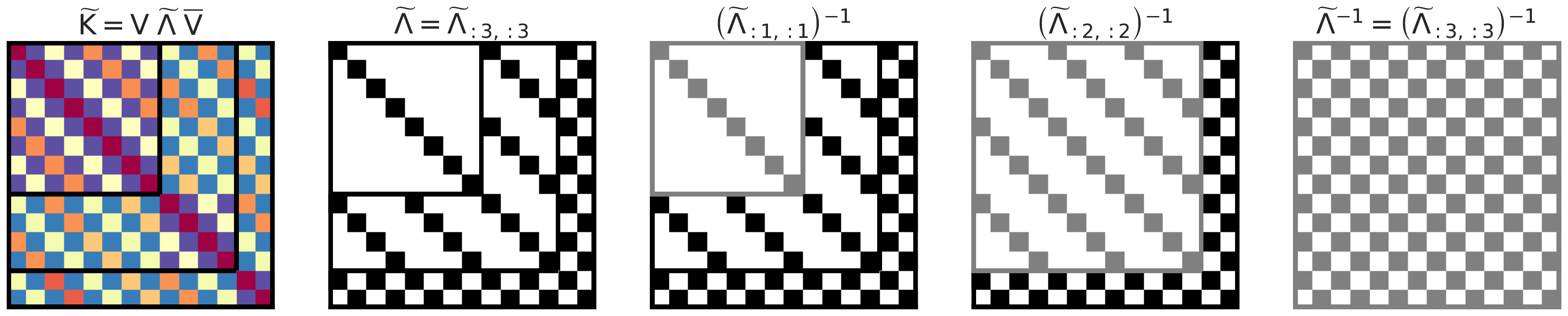}
    \caption{Structure and the inversion algorithm for a $T=3$ fast MTGP with a shifted lattice (after bit-reversal permutation) and a shift-invariant kernel.}
    \label{fig:kmat_inv}
    \vspace{-.75cm}
\end{figure}

To motivate our fast inverse and determinant algorithm, let us consider an example with $T=3$ tasks having $n_1 = 8$, $n_2 = 4$ and $n_3 = 2$. \Cref{fig:kmat_inv} visualizes the resulting Gram matrix $\tmK$ (in the DSI-DSEQ setting), the sparsity pattern of $\tmLambda$, and the $T=3$ stage inversion algorithm. In the first stage, we compute $\tmLambda_{11}^{-1}$ which is simply the inverse of a diagonal matrix. In the second stage we compute 
\begin{align*}
    \tmLambda_{:2,:2}^{-1} = 
    \begin{bsmallmatrix}
        \tmLambda_{11}^{-1} + \tmLambda_{11}^{-1} \tmLambda_{12} \mS_2^{-1} \tmLambda_{21} \tmLambda_{11}^{-1} & -\tmLambda_{11}^{-1} \tmLambda_{12} \mS_2^{-1} \\
        - \mS_2^{-1} \tmLambda_{21} \tmLambda_{11}^{-1} & \mS_2^{-1}
    \end{bsmallmatrix} 
\end{align*}
where $\mS_2 = \tmLambda_{22} - \tmLambda_{21} \tmLambda_{11}^{-1} \tmLambda_{12}$ is the Schur complement. We also have that $\lvert \tmLambda_{:2,:2} \rvert = \lvert \tmLambda_{11} \rvert \cdot \lvert \mS_2 \rvert$. In the third and final stage we compute
\begin{align*}
    \tmLambda^{-1} = \tmLambda_{:3,:3}^{-1}
    &= \begin{bsmallmatrix}
        \tmLambda_{:2,:2} & \tmLambda_{:2,3} \\
        \tmLambda_{3,:2} & \tmLambda_{33}
    \end{bsmallmatrix}^{-1}
    = 
    \begin{bsmallmatrix}
        \tmLambda_{:2,:2}^{-1} + \tmLambda_{:2,:2}^{-1} \tmLambda_{:2,3} \mS_3^{-1} \tmLambda_{3,:2} \tmLambda_{:2,:2}^{-1} & -\tmLambda_{:2,:2}^{-1} \tmLambda_{:2,3} \mS_3^{-1} \\
        -\mS_3^{-1} \tmLambda_{3,:2} \tmLambda_{:2,:2}^{-1} & \mS_3^{-1}
    \end{bsmallmatrix}
\end{align*}
where $\mS_3 = \tmLambda_{33} - \tmLambda_{3,:2} \tmLambda_{:2,:2}^{-1} \tmLambda_{:2,3}$ is another Schur complement and $\lvert \tmLambda \rvert = \lvert \tmLambda_{:3,:3} \rvert = \lvert \tmLambda_{:2,:2} \rvert \cdot \lvert S_3 \rvert$. Crucially, the Schur complements $\mS_t$ are all diagonal and $\tmLambda^{-1}_{:t,:t}$ only depends on $\tmLambda^{-1}_{:t-1,:t-1}$, $\tmLambda_{:t-1,t}$, and $\tmLambda_{tt}$. \Cref{algo:inv_det_Theta} generalizes the above routine for computing the inverse $\tmLambda^{-1}$ and determinant $\lvert \tmLambda \rvert$ to any number of tasks $T \in \bbN$. 

\subsection{Complexity Analysis}

We now give a complexity analysis of the computational and storage costs for computing the matrix inverse and determinant of $\tmK$ using \Cref{algo:inv_det_Theta} following the comments therein. Note that the inverse and determinant of $\tmLambda$ are well defined since $\tmLambda_{:t,:t}$ being HPD for each $t \in \{1,\dots,T\}$ implies that each Schur complement $\mS_t$ is also HPD and hence invertible with $\llvert \mS_t \rrvert > 0$. \Cref{thm:costs_fast_mtgp} provides this analysis for the general case of unequal sample sizes over tasks. \Cref{corr:costs_fast_mtgp_equal} gives this analysis for the specific case of equal sample sizes. These cost and storage complexities are simplified in \Cref{table:costs_storage} under the assumption that $d = \calO(1)$ and $T = \calO(1)$. The following theorem was proven by accumulating the costs in \Cref{algo:inv_det_Theta}.

\begin{algorithm}[!t]
    \fontsize{10}{10}\selectfont
    \caption{Inverse and log-determinant of $\tmLambda$}
    \label{algo:inv_det_Theta}
    \begin{algorithmic}
        \Require $n_1 \geq \dots \geq n_T$ with $n_t = 2^{m_t}$ and $\tmLambda$ as in \Cref{thm:FMTGP_computations}.
        \State $\mD \gets \tmLambda_{11}$ \Comment{A $2^{m_1} \times 2^{m_1}$ diagonal matrix.}
        \State $\mA \gets \mD^{-1}$ \Comment{A $2^{m_1} \times 2^{m_1}$ diagonal matrix. Costs $\calO(2^{m_1})$.} 
        \State $\varrho \gets  \log \lvert \mD \rvert$ \Comment{A scalar. Costs $\calO(2^{m_1})$.}
        \State $t \gets 2$
        \While{$t \leq T$}
            \State $\mD \gets \tmLambda_{tt}$ \Comment{A $2^{m_t} \times 2^{m_t}$ diagonal matrix.}
            \State $\mB \gets \tmLambda_{:(t-1),t}$ \Comment{A $\sum_{t'=1}^{t-1} 2^{m_{t'} - m_t} \times 1$ block matrix with $2^{m_t} \times 2^{m_t}$ diagonal blocks.}
            \State $\mL \gets \mA \mB$ \Comment{A $\sum_{t'=1}^{t-1} 2^{m_{t'}-m_{t}} \times 1$ block matrix with $2^{m_t} \times 2^{m_t}$ diagonal blocks. Costs $\calO([\sum_{t'=1}^{t-1} 2^{m_{t'} - m_{t}}][\sum_{t'=1}^{t-1} 2^{m_{t'}}])$.}
            \State $\mF \gets \mB^\dagger \mL$ \Comment{A $2^{m_t} \times 2^{m_t}$ diagonal matrix. Costs $\calO(\sum_{t'=1}^{t-1} 2^{m_{t'}})$.}
            \State $\mS \gets \mD-\mF$ \Comment{A $2^{m_t} \times 2^{m_t}$ diagonal matrix (the Schur complement). Costs $\calO(2^{m_t})$.}
            \State $\varrho \gets \varrho + \log \lvert \mS \rvert$ \Comment{A scalar. Costs $\calO(2^{m_t})$. Equivalent to $\log \lvert \tmLambda_{:t,:t} \rvert$.}
            \State $\mG \gets \mS^{-1}$ \Comment{A $2^{m_t} \times 2^{m_t}$ diagonal matrix. Costs $\calO(2^{m_t})$.}
            \State $\mH \gets \mL \mG$ \Comment{A $\sum_{t'=1}^{t-1} 2^{m_{t'}-m_{t}} \times 1$ block matrix with $2^{m_t} \times 2^{m_t}$ diagonal blocks. Costs $\calO(\sum_{t'=1}^{t-1} 2^{m_{t'}})$.}
            \State $\mJ \gets \mH \mL^\dagger$ \Comment{A $\sum_{t'=1}^{t-1} 2^{m_{t'}-m_{t}} \times \sum_{t'=1}^{t-1} 2^{m_{t'}-m_{t}}$ block matrix with $2^{m_t} \times 2^{m_t}$ diagonal blocks. Costs $\calO([\sum_{t'=1}^{t-1} 2^{m_{t'} - m_t}][\sum_{t'=1}^{t-1} 2^{m_{t'}}])$.}
            \State $\mM \gets \mA + \mJ$ \Comment{A $\sum_{t'=1}^{t-1} 2^{m_{t'}-m_{t}} \times \sum_{t'=1}^{t-1} 2^{m_{t'}-m_{t}}$ block matrix with $2^{m_t} \times 2^{m_t}$ diagonal blocks. Costs $\calO([\sum_{t'=1}^{t-1} 2^{m_{t'} - m_t}][\sum_{t'=1}^{t-1} 2^{m_{t'}}])$.}
            \State $\mA \gets \begin{bsmallmatrix} \mM & -\mH \\ -\mH^\dagger & \mG \end{bsmallmatrix}$ \Comment{A $\sum_{t'=1}^t 2^{m_{t'}-m_t} \times \sum_{t'=1}^t 2^{m_{t'}-m_t}$ block matrix with $2^{m_t} \times 2^{m_t}$ diagonal blocks.}
            \State $t \gets t+1$
        \EndWhile
        \State $\tmLambda^{-1} \gets \mA$ \Comment{A $\sum_{t=1}^T 2^{m_{t}-m_T} \times \sum_{t=1}^T 2^{m_{t}-m_T}$ block matrix with $2^{m_T} \times 2^{m_T}$ diagonal blocks.}
        \State $\log \lvert \tmLambda \rvert \gets \varrho$ \Comment{A scalar.}
        \\\Return $\tmLambda^{-1}, \log \lvert \tmLambda \rvert$
    \end{algorithmic}
\end{algorithm}

\begin{theorem} \label{thm:costs_fast_mtgp}
    Suppose \Cref{cond:FMTGP} holds, crucially with $n_1 \geq \cdots \geq n_T$. Then evaluating $\tmLambda \in \bbC^{N \times N}$ requires $\calO\left(\sum_{t=1}^T (T-t+1) (n_t \log n_t + d n_t)\right)$ computations and \patchoverfullhbox $\calO\left(\sum_{t=1}^T (T-t+1) n_t\right)$ storage. Having evaluated $\tmLambda$, computing $\tmK \by = \mV \tmLambda \, \mV^\dagger \by$ for any $\by \in \bbR^N$ requires $\calO\left(\sum_{t=1}^T n_t \log n_t + \sum_{t=1}^T (T-t+1) n_t\right)$ computations. Having evaluated $\tmLambda$, evaluating $\lvert \tmK \rvert = \lvert \tmLambda \rvert$ and $\tmLambda^{-1}$ using \Cref{algo:inv_det_Theta} requires $\calO\left(\sum_{t=2}^T \left(\sum_{t'=1}^{t-1} n_{t'}\right)^2/n_t\right)$ computations and $\calO(N^2/n_T)$ storage. Having stored $\tmLambda^{-1}$, evaluating $\tmK^{-1} \by= \mV \tmLambda^{-1} \mV^\dagger \by$ for any $\by \in \bbR^N$ requires $\calO\left(\sum_{t=1}^T n_t \log n_t + N^2/n_T\right)$ computations. In all, fast MTGPs fitting requires $\calO\left(\sum_{t=1}^T (T-t+1)(n_t \log n_t + d n_t) + \sum_{t=2}^T \left(\sum_{t'=1}^{t-1} n_{t'}\right)^2/n_t + N^2/n_T\right)$ computations and $\calO\left(\sum_{t=1}^T (T-t+1)n_t + N^2/n_T\right)$ storage.
\end{theorem}

Let us consider the common setting of a relatively small dimension $d = \calO(1)$ and number of tasks $T = \calO(1)$. In such a case, the summarized requirements in \Cref{table:costs_storage} imply that standard MTGPs require $\calO(N^3)$ computations and $\calO(N^2)$ storage while fast MTGPs require only $\calO(N \log N + N^2/n_\mathrm{min})$ computations and $\calO(N^2/n_\mathrm{min})$ storage where $n_\mathrm{min} = \min\{n_1,\dots,n_T\}$. Therefore, with a fixed budget $N$, fast MTGPs become more efficient when the minimum sample size $n_\mathrm{min}$ grows or when the sample sizes for each task $n_1,\dots,n_T$ grow closer together. More specifically, $NT \lesssim N^2/n_\mathrm{min} \lesssim N^2$ where the lower bound is attained when all $n_t$ are equal, i.e., $n_1=\cdots=n_T = N/T$, and the upper bound is attained when $n_\mathrm{min}=1$, i.e., taking only one sample for one of the tasks. In the worst case, fast MTGPs require the same $\calO(N^2)$ storage as standard MTGPs while reducing the computational burden from $\calO(N^3)$ to $\calO(N^2)$. In the best case, fast MTGPs only require $\calO(N)$ storage and $\calO(N \log N)$ computations. 
In the optimal case of equal sample sizes we have the following corollary of \Cref{thm:costs_fast_mtgp}.

\begin{corollary} \label{corr:costs_fast_mtgp_equal}
    Suppose \Cref{cond:FMTGP} holds for $n:=n_t = 2^{m}$ with $m \in \bbN_0$ for all $t \in \{1,\dots,T\}$, so $N=Tn$. Evaluating $\tmLambda \in \bbC^{N \times N}$ requires $\calO\left(T^2(n \log n + dn)\right)$ computations and $\calO\left(T^2 n\right)$ storage. Having evaluated $\tmLambda$, computing $\tmK \by = \mV \tmLambda \, \mV^\dagger \by$ for any $\by \in \bbR^N$ requires $\calO\left(T n \log n + T^2 n\right)$ computations. Having evaluated $\tmLambda$, evaluating $\lvert \tmK \rvert = \lvert \tmLambda \rvert$ and $\tmLambda^{-1}$ using \Cref{algo:inv_det_Theta} requires $\calO\left(T^3 n\right)$ computations and $\calO(T^2n)$ storage.  Having stored $\tmLambda^{-1}$, evaluating $\tmK^{-1} \by = \mV \tmLambda^{-1} \mV^\dagger \by$ for any $\by \in \bbR^N$ requires $\calO\left(T n \log n + T^2 n\right)$ computations. In all, fast MTGPs with equal sample sizes for each task require $\calO\left((n \log n)T^2 + n T^3 + nT^2d\right)$ computations and 
    $\calO\left(nT^2\right)$ storage.
\end{corollary}



\subsection{Extensions for Bayesian Cubature}

While the focus of this paper is on \textit{predicting} the black-box function $f$, there is also a growing body of work on Bayesian cubature, which makes use of GP surrogates for \textit{integral} approximations (see, e.g., \cite{o1991bayes,briol2019probabilistic}). Bayesian cubature has been adopted in broad applications, including uncertainty quantification
\cite{diaconis.bayesian_numerical_analysis,rathinavel.bayesian_QMC_lattice,rathinavel.bayesian_QMC_sobol,sorokin.FastBayesianMLQMC}, 
multilevel (quasi-)Monte Carlo 
\cite{giles.MLMC_path_simulation,robbe.multi_index_qmc,sorokin.FastBayesianMLQMC},
and PDE-constrained inverse problems in probabilistic numerics 
\cite{cockayne2017probabilistic}. 

Bayesian cubature aims to compute the integral quantity $\mu := \int_\calX f(\bx) \D \bx$. Under a (single-task) GP surrogate on $f$, one can show \cite{o1991bayes} that the posterior distribution of the integral $\mu$ takes the form $\mu \vert \mX, \by \sim \calN(\hmu,\sigma^2)$ where $\hmu := \int_\calX \bbE[f(\bx) | \mX,\by] \D \bx$ serves as the point estimate of the integral, and $\sigma^2 := \int_\calX \int_\calX \Cov[f(\bx),f(\bx') | \mX] \D \bx \D \bx'$ provides its uncertainty. The fast single-task GP Bayesian cubature form is giving in the following theorem, as proven in \cite{rathinavel.bayesian_QMC_lattice,rathinavel.bayesian_QMC_sobol}. 

\begin{theorem}[Fast Bayesian cubature] \label{thm:FBC}
    Suppose \Cref{cond:FGP} holds and that for some $\gamma \in \bbR_+$ we have that $\int_{[0,1]^d} K(\bx,\bx') \D \bx' = \gamma$ for all $\bx \in [0,1]^d$. Let  $\tau = \htau_\mathrm{MLE} = \bone^\intercal \by / n$ or $\tau = \htau_\mathrm{GCV} = \bone^\intercal \by / n$. Then $\hmu = \bone^\intercal \by / n$ and $\sigma^2 = \gamma - \gamma^2 (\bone^\intercal \tmK_{:,0}/n)^{-1}$.
\end{theorem}

Multitask Bayesian cubature for the mean vector $\bmu = \left(\int_\calX f(t,\bx) \D \bx\right)_{t=1}^T$ has the conditional distribution $\bmu \vert \mX, \by \sim \calN(\hbmu,\mSigma)$. We are often interested in a linear combination of integrals for each task. Suppose we have user-provided weights $\bchi \in \bbR^T$ for which our quantity of interest is $\bchi^\intercal \bmu \sim \calN(\bchi^\intercal \hbmu,\bchi^\intercal \mSigma \bchi)$, e.g., $\bchi = (0,\dots,0,1)^\intercal$ to select the final task mean or $\bchi = \bone$ if $f(t,\bx)$ models the difference of consecutive tasks as in multilevel Monte Carlo \cite{giles.MLMC_path_simulation,robbe.multi_index_qmc,sorokin.FastBayesianMLQMC}. The fast MTGP Bayesian cubature form is given in the following theorem. The SI and DSI product kernels in \eqref{eq:SI_kernel_1d_ex} and \eqref{eq:DSI_kernel_1d_ex} respectively satisfy the constant-mean conditions in both theorems.

\begin{theorem}
    Suppose \Cref{cond:FMTGP} holds and there is some $\gamma \in \bbR_+$ such that \patchoverfullhbox $\int_{[0,1]^d} Q(\bx,\bx') \D \bx' = \gamma$ for all $\bx \in [0,1]^d$. Then $\hbmu = \btau + \gamma \mR \mE \tmK^{-1}(\by-\bE \btau)$ and \patchoverfullhbox $\mSigma = \gamma \mR - \gamma^2 \mR \mPi \mR \in \bbR^{T \times T}$. If $\btau = \hbtau_\mathrm{MLE}$ then $\hbmu = \hbtau_\mathrm{MLE}$.
\end{theorem}


\section{Numerical Experiments} \label{sec:numerics}

\subsection{Setting}

Our experiments make use of the \texttt{FastGPs} package \patchoverfullhbox (\url{https://alegresor.github.io/fastgps/}) \cite{sorokin.fastgps_probnum25,sorokin.FastBayesianMLQMC} with the linked documentation containing a demo reproducing the numerical experiments from this paper. \texttt{FastGPs} uses the quasi-Monte Carlo package \texttt{QMCPy} (\url{https://qmcsoftware.github.io/QMCSoftware/}) \cite{sorokin.qmcpy_joss} for generating low-discrepancy lattices and digital sequences. We will compare the following model fitting approaches, where each approach is implemented for both the multitask GP setting and the single-task GP setting (where separate GPs are trained for each task).

\begin{itemize}[leftmargin=*]
    \item \textbf{Cholesky SE}: This is the standard approach for fitting a GP (or MTGP), where a full Cholesky decomposition is used for the matrix inversion and determinant computation of the Gram matrix $\tmK$. Here, we use the \texttt{FastGPs} implementation with the squared exponential (Gaussian) input kernel.
    \item \textbf{CG SE (\texttt{GPyTorch})}: This is the (preconditioned) conjugate gradient (CG) benchmark, which is widely used for approximating the Gram matrix calculations for GP (and MTGP) training \cite{gardner.gpytorch_GPU_conjugate_gradient}. In particular, this approach leverages the Lanczos tridiagonalization algorithm and stochastic trace estimation to approximate the log-determinant and its gradient, respectively. Here, we use the \texttt{GPyTorch} \cite{gardner.gpytorch_GPU_conjugate_gradient} implementation with the squared exponential input kernel.
    \item \textbf{Fast DSI (proposed)}: This approach makes use of the design-kernel pairings discussed in our paper. The single-task GP setting follows the framework in \cite{rathinavel.bayesian_QMC_lattice,rathinavel.bayesian_QMC_sobol} (see also the later parts of \Cref{sec:methods}), and the multitask GP setting is our proposed approach in \Cref{sec:MTGPs}. Here, we use the \texttt{FastGPs} implementation with digitally-shift-invariant kernels paired with digitally-shifted design points. The pairing with shift-invariant kernels is not tested here, since the test functions in our experiment are not necessarily periodic in nature.
\end{itemize}

\begin{wraptable}[11]{l}{0.5\textwidth}
    \centering
    \resizebox{0.5\textwidth}{!}{%
    \begin{tabular}{rccc}
        problem & $d$ & $T$ & $\bn \in$ \\
        \toprule 
        Rosenbrock & $2$ & $3$ & $\{\{2^{15},2^{14},2^{13}\}\}$ \\
        Ackley & $4$ & $2$ & $\times_{t=1}^T \{2^{15},2^{14},\dots,2^5\}$ \\ 
        borehole & $8$ & $2$ & $\times_{t=1}^T \{2^{15},2^{14},\dots,2^5\}$ \\
        elliptic PDE & $16$ & $3$ & $\times_{t=1}^T \{2^{13},2^{11},2^9,2^7,2^5\}$ \\ 
        cookie & $8$ & $4$ & $\times_{t=1}^T \{2^{13},2^{11},2^9,2^7,2^5\}$ \\
        \bottomrule
    \end{tabular}
    }
    \caption{Problem parameters including the dimension $d$, the number of tasks $T$, and sample sizes $\bn$ considered.}
    \label{table:problem_parameters}
\end{wraptable}

All methods make use of the same digitally-shifted design points (see \Cref{def:dseqs}) with separate random shifts for each task, which gives different design points for different tasks. Different sample sizes are used for different tasks, following the setting in \Cref{table:problem_parameters}. For all approaches, model parameters are fitted via MLE, by minimizing the loss function \eqref{eq:MTGP_MLE_GCV_loss} using the resilient backpropagation algorithm \cite{riedmiller1993direct} (\texttt{Rprop} in \texttt{PyTorch}) .
All computations were run on an M3 MacBook Pro in double precision. 

\subsection{Benchmarks} \label{sec:results_benchmark}

\Cref{table:problem_parameters} outlines the setting for each benchmark problem in the following suite of experiments. 

\begin{itemize}[leftmargin=*]
    \item \textbf{Rosenbrock.} The $d=2$ Rosenbrock function with $T=3$ variants of this function in \cite{wackers2023efficient}: $f(1,\bx) = (f(3,\bx)-4-0.5 x_1-0.5 x_2)/(10+0.25 x_1+0.25 x_2)$, $f(2,\bx) = 50(x_2-x_1^2)^2+(-2-x_1)^2-80-0.25 x_1x_2$, and $f(3,\bx) = 100(x_2-x_1^2)^2+(1-x_1)^2.$ 
    \item \textbf{Ackley.} The Ackley function is $f(\bx) = g(\bPhi^{-1}(\bx))$ where $\bPhi^{-1}(\bx) = 65.536\bx-32.768$ for $g(\bz) = -20 \exp\left(-0.2 \sqrt{\tfrac{1}{d}\sum_{j=1}^d z_j^2}\right) - \exp\left(\tfrac{1}{d} \sum_{j=1}^d \cos(c z_j)\right) + 20 + \exp(1)$ from the Virtual Library of Simulation Experiments (VLSE) \cite{VLSE}. Task $t=1$ sets $c=0$ while task $t=2$ sets $c = 2\pi$. For $\bX \sim \calU[0,1]^d$ we have $\bZ = \bPhi^{-1}(\bX) \sim \calU[-32.768,32.768]^d$. We will consider the $d=4$ case. 
    \item \textbf{Borehole.} The borehole function models water flow through a borehole with $f(\bx) = g(\bPhi^{-1}(\bx))$ where $g(\bz) = c_1 \pi T_u (H_u-H_l) \left(c_2+\tfrac{2 L T_u}{\log(r/r_w)r_w^2 K_w}+\tfrac{T_u}{T_l}\right)^{-1} / \log(r/r_w)$. Here $\bPhi^{-1}$ applies the inverse CDF transform so that for $\bX \sim \calU[0,1]^d$ we have that \patchoverfullhbox $(r_w,r,T_u,H_u,T_l,H_l,L,K_w) = \bPhi^{-1}(\bX)$ with independent marginals specified by the VLSE \cite{VLSE}. Following \cite{xiong2013sequential}, task $t=1$ sets $(c_1,c_2) = (2,1)$ while task $t=2$ sets $(c_1,c_2) = (5,3/2)$.
    \item \textbf{Elliptic PDE.} Let us consider the one-dimensional elliptic PDE $-\nabla(e^{a(u,\bx)} \nabla F(u,\bx)) = g(u)$ with $u \in [0,1]$ and boundary conditions $F(0,\bx) = F(1,\bx) = 0$ for all $\bx \in [0,1]^d$. The forcing term $g$ is set to the constant $g(u) = 1$ for all $u \in [0,1]$. We will generate $a(u,\bx) = \sum_{j=1}^d \Phi^{-1}(x_j) \sin(\pi k u)/j$ in $d=16$ dimensions and $\Phi$ the CDF of the standard normal so $\Phi^{-1}(X_j) \sim \calN(0,1)$. Let us denote by $F_t$ the task $t$ numerical PDE solution using a finite difference method with $2^{1+t}+1$ evenly spaced mesh points $(k/2^{1+t})_{k=0}^{2^{1+t}}$. We will take the discretized solution to be $f(t,\bx) = \max_u F_t(u,\bx)$ and use $T=3$ tasks. For each query of $f(t,\cdot)$ we need to solve a tridiagonal linear system, which has linear complexity.
    \item \textbf{Cookies in the oven.} This problem simulates baking cookies subject to uncertainty in the conductivity coefficients in $d=8$ circular subdomains, the cookies. 
    For task $t \in \{1,2,3,4\}$ we used $100t$ mesh points within the finite element solver in each dimension, e.g., for our two-dimensional physical domain the $t = 4$ mesh has $400^2$ nodes. We again take $f(\bx) = g(\bPhi^{-1}(\bX))$ where $g$ is the numerical solver accessed through \texttt{UM-Bridge} \cite{umbridge.software} and $\bPhi^{-1}(\bx) = 0.79\bx - 0.99$ so for $\bX \sim \calU[0,1]^d$ we have $\bPhi^{-1}(\bX) \sim \calU[-0.99,-0.2]$. 
\end{itemize}

\begin{figure}[!t]
    \centering 
     \includegraphics[width=1\textwidth]{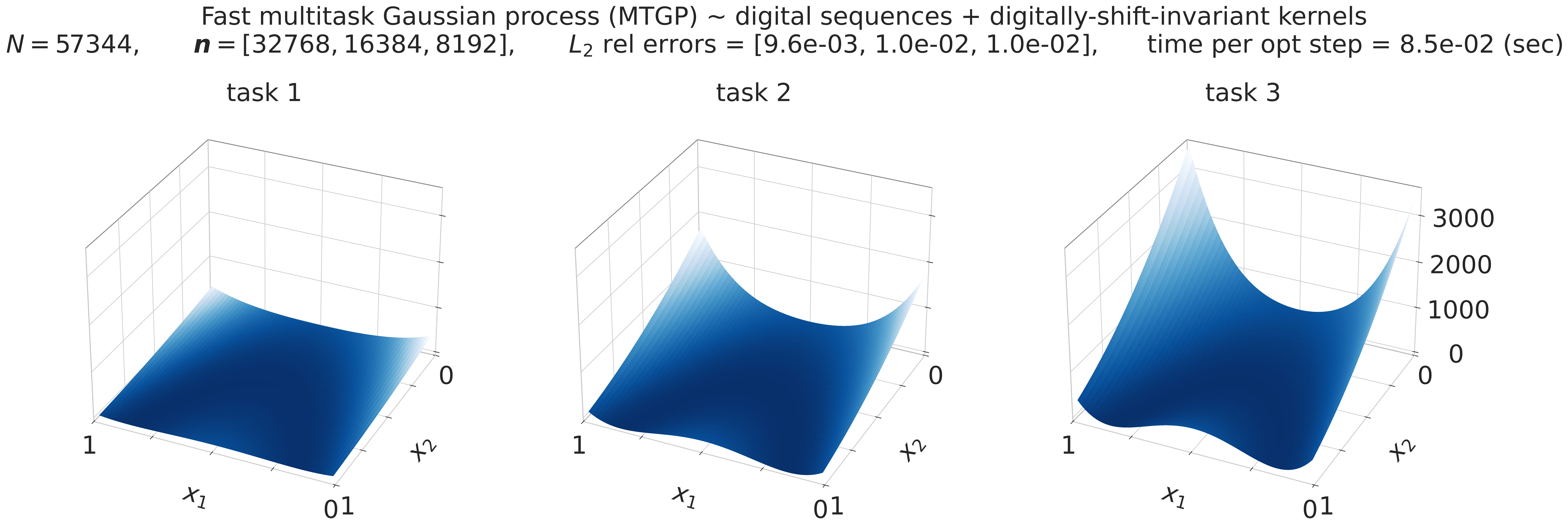}
    \caption{Posterior mean estimates of a fast MTGP fit to the multitask Rosenbrock function.}
    \label{fig:rosenbrock}
    \vspace{-.75cm}
\end{figure}

\Cref{fig:rosenbrock} visualizes our fast MTGP DSI-DSEQ variant fit to the Rosenbrock functions with $N = 57344$ total points. Equivalent fitting for standard MTGPs would require large scale computing resources and be infeasible on a personal laptop. Fitting our fast MTGP for $217$ MLE optimization steps took less than 15 seconds on a laptop with each optimization step requiring less than a tenth of a second. Moreover, the multitask DSI kernels were accurate enough to recover solutions at each task to less than $1\%$ $L^2$ relative error.

For the Ackley, borehole, elliptic PDE, and cookie problems we ran experiments across a wide range of $\bn$ values in grids specified in \Cref{table:problem_parameters}. For each $\bn$, we ran $5$ independent trails and compare the median time per optimization step against both the median $L^2$ relative prediction test errors and Bayesian cubature errors on the final task. $L^2$ relative errors are approximated using $1024$ LD test points. For the CG SE variant, Bayesian cubature is not directly supported in \texttt{GPyTorch}, so we use the $1024$ LD test points to compute a QMC estimate of the Bayesian cubature mean. Due to memory and time constraints, we only run Cholesky SE when $N \leq 2048$, and we only ran CG SE (\texttt{GPyTorch}) when $N \leq 4096$. All benchmark experiments used $100$ GP hyperparameter optimization steps.

\begin{figure}[!t]
    \centering 
    \includegraphics[width=.9\textwidth]{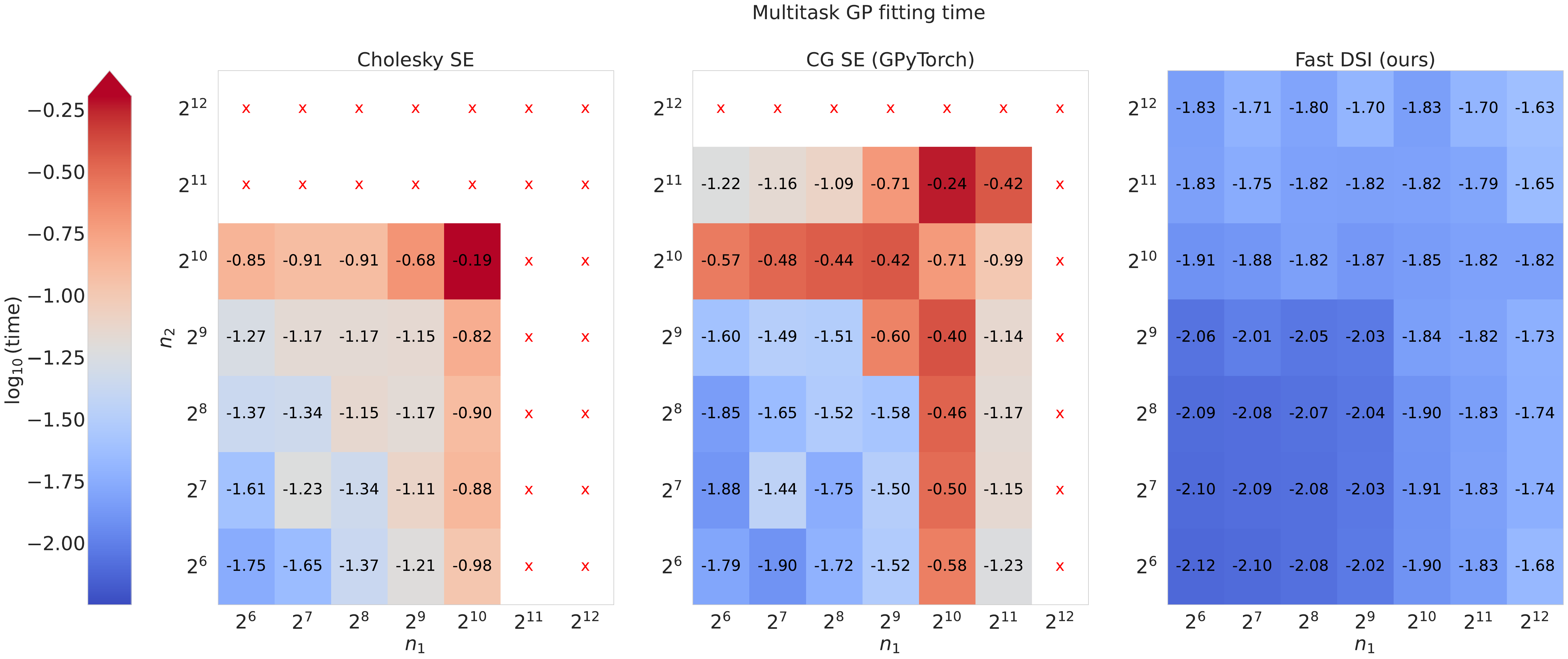}
    \includegraphics[width=.9\textwidth]{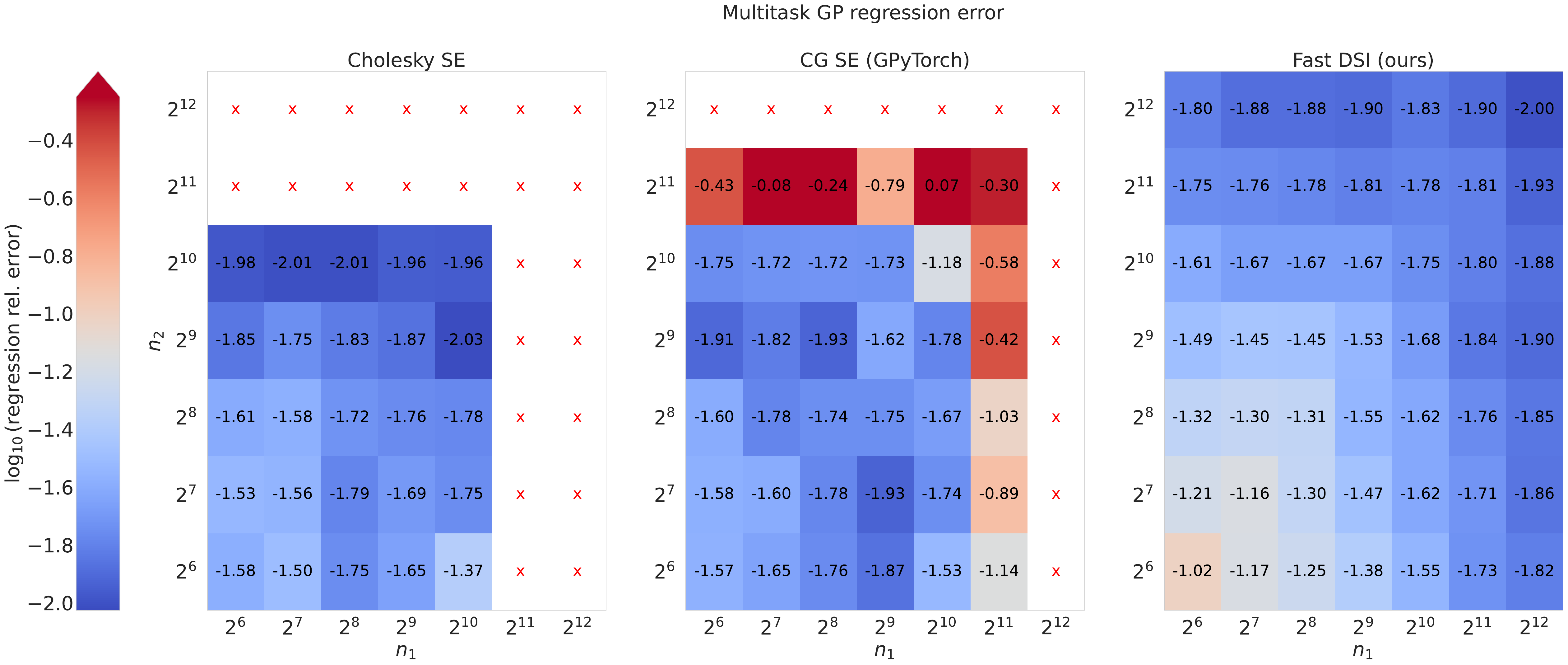}
    \caption{Borehole problem analytics. The top row shows time per optimization step (seconds). The bottom row shows the regression $L^2$ relative errors.}
    \label{fig:borehole_times_errors}
    \vspace{-1cm}
\end{figure} 


Overall, the benchmark results show that out fast DSI MTGP construction significantly reduces the required fitting time and memory requirements while achieving comparable errors to the full Cholesky SE and CG SE (\texttt{GPyTorch}) methods. We also see that fast MTGPs provide improved accuracy compared fast single-task GPs with comparable sample sizes. Fast MTGPs fitting is able to scale up to large datasets, over 50 thousand samples as tested, in just a few seconds on a personal computer. The predominant existing MTGP methods we compare against are generally unable to accommodate sample sizes beyond a few thousand without exhausting memory and compute resources or sacrificing accuracy to approximate MTGP constructions. 

We found the CG SE results from the \texttt{GPyTorch} implementation to be quite volatile in terms of both timing and accuracy due to their automated choices for the number of PCG steps and PCG error tolerances. For small sample sizes or nearly independent tasks, CG SE did provide comparable accuracy to Cholesky SE and our fast DSI MTGPs while being slightly faster than Cholesky SE, but still significantly slower than our fast DSI MTGPs. However, in the regime of interest with large sample sizes and correlated tasks, the highly ill-conditioned Gram matrices lead CG SE to have poor PCG performance resulting in subpar accuracy compared to both the standard Cholesky SE and our fast DSI MTGPs. Going forward, we will mainly focus our error comparisons between Cholesky SE and fast DSI MTGPs. 

\begin{figure}[!t]
    \centering 
    \includegraphics[width=1\textwidth]{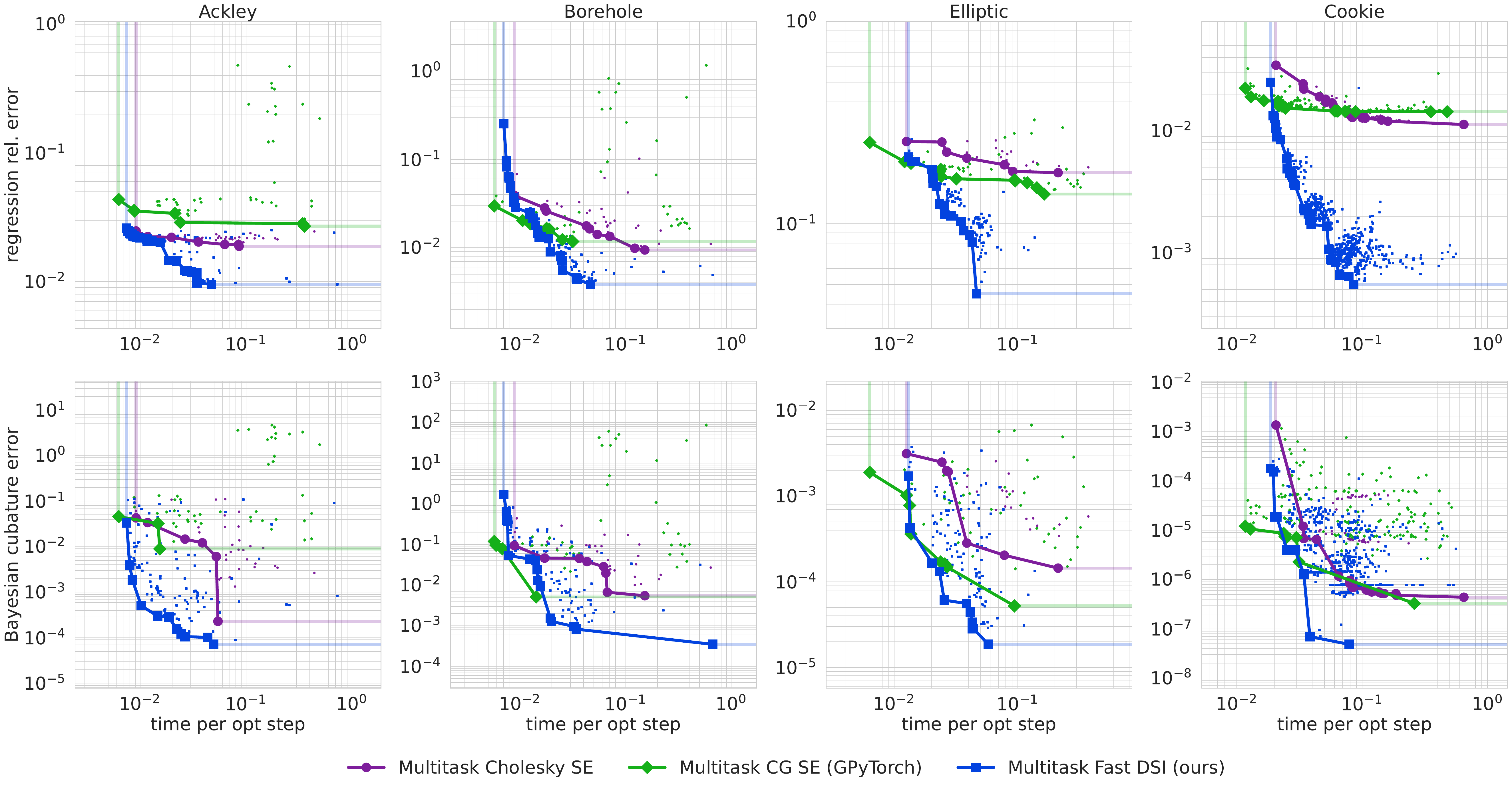}
    \caption{Pareto frontiers for $L^2$ relative regression errors and Bayesian cubature errors versus time per optimization step (seconds).}
    \label{fig:errors}
    \vspace{-1cm}
\end{figure} 

\begin{figure}[!t]
    \centering 
    \includegraphics[width=1\textwidth]{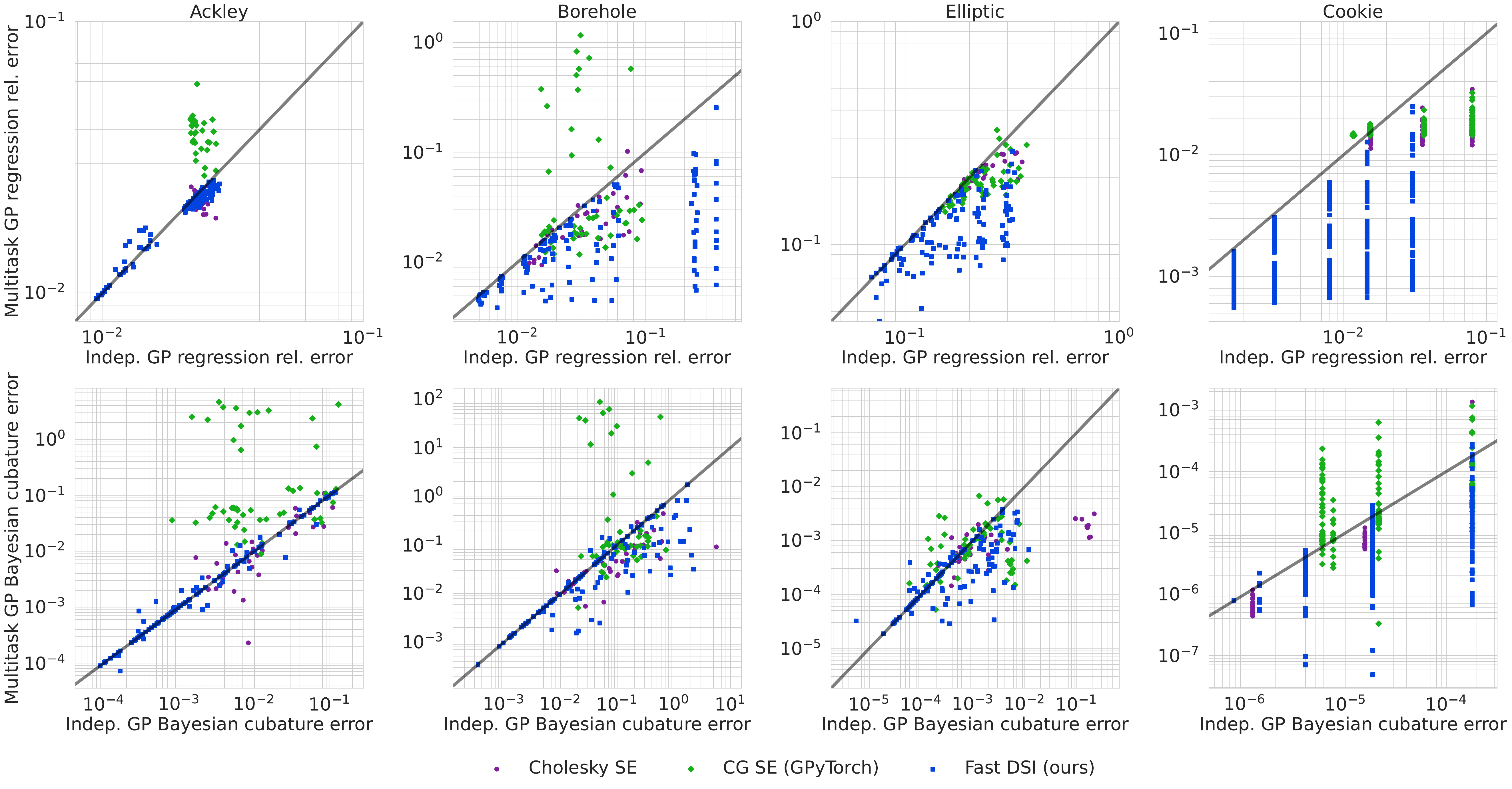}
    \caption{$L^2$ relative regression errors and Bayesian cubature errors compared across single-task and multitask GPs.}
    \label{fig:errors_st_vs_mt}
    \vspace{-1cm}
\end{figure}

\Cref{fig:errors_st_vs_mt} shows all MTGPs methods outperform their single-task counterparts across our suite of benchmarks. Specifically, this verifies our fast DSI MTGPs are capable of exploiting correlated task data to enhance the multitask surrogate. As expected, the multitask modeling advantage grows as the sample sizes skew towards the final task of interest or tasks highly-correlated to the final tasks.

The borehole fitting-time analysis in the top row of \Cref{fig:borehole_times_errors} shows that our fast DSI MTGPs required  less than 0.025 seconds per optimization step across all plotted sample sizes $n_1,n_2 \in \{2^6,2^7,\dots,2^{12}\}$, including the maximum sample setting with $N=2^{13}=8192$ total samples across both tasks. In contrast, full Cholesky SE GP required over 0.025 seconds per optimization step whenever the total sample size $N=n_1+n_2$ exceeded $256$, and required over 0.5 seconds per step on the largest tested Cholesky SE setting with $n_1=n_2=2^{10}$ giving $N=2048$. In this setting, our fast DSI MTGPs fitting was over $45$ times faster than the Cholesky SE construction. Compared to CG SE, our fast DSI MTGP construction was over $23$ times faster at the largest comparable setting of $n_1=n_2=2^{11}$ giving $N=4096$.

For the borehole accuracy analysis in \Cref{fig:borehole_times_errors} we found that our fast DSI MTGPs provided comparable errors to the Cholesky SE CG SE constructions. At the same sample sizes $\bn$, the SE Cholesky constructions yielded slightly improved accuracy compared to the fast DSI MTGPs due to the smoothness in the borehole problem being better exploited by the SE kernel compared to the DSI kernel. However, when pushing fast DSI MTGPs to larger $\bn$ outside the feasible region for existing methods, we observed the lowest errors achieved across all three methods. 
We now study the accuracy versus time per optimization step tradeoff across the full suite of benchmarks. 

For the full benchmark suite, \Cref{fig:errors} compares the time per optimization step against both the $L^2$ relative regression errors and the absolute Bayesian cubature errors. The approximate Pareto frontier is plotted along which improved errors can only be observed by increasing the time per optimization step. Here we see the fast DSI MTGPs provide a substantially improved tradeoff between fitting time and accuracy compared to existing methods. For a comparable error tolerance, the fast DSI MTGPs required only slightly more samples than existing methods while significantly reducing fitting time. For a comparable fitting time budget, the fast DSI MTGPs are able to accommodate orders of magnitude more samples than existing methods which enable significantly lowers errors. For instance, in the cookie problem at around $0.1$ seconds per optimization step, fast DSI MTGPs were able to achieve $L^2$ relative regression errors of nearly $0.1\%$ whereas Cholesky SE and CG SE methods achieved errors of only $1\%$. In the same setting, fast DSI MTGPs have Bayesian cubature errors around $5 \times 10^{-8}$ while Cholesky SE and CG SE methods had errors around $10^{-6}$. For the elliptic PDE problem the fast DSI MTGPs achieved $L^2$ regression errors around $5\%$ with around $0.05$ seconds per optimization step, whereas Cholesky SE and CG SE methods had errors over $10\%$ error even at the largest budgets of over $0.3$ seconds per optimization step. In the same setting, the fast DSI constructions provided an order of magnitude savings in $L^2$ relative error compared to Cholesky SE MTGPs. 

\subsection{Surrogate Modeling of the Quark-Gluon Plasma} 

Finally, we investigate the effectiveness of the proposed fast MTGP in an application on the surrogate modeling of the quark-gluon plasma \cite{cao2021determining,everett2021multisystem}. The QGP is an exotic state of nuclear matter that once filled the universe shortly after the Big Bang, and the study of this plasma sheds light on the origins of matter. This plasma can be produced at a small scale by colliding together atomic nuclei at near-light speeds \cite{liyanage2022efficient} within particle accelerators (e.g., at CERN), during which the extreme temperature and pressure converts the colliding particles into a plasma of subatomic particles, namely the QGP.

\begin{figure}[!t]
    \centering
    \begin{minipage}{0.50\linewidth}
        \centering
        \small
        \resizebox{\textwidth}{!}{%
        \begin{tabular}{ll p{5cm}}
        \toprule
        Parameter & Range & Description \\
        \hline
        $B_G[\mathrm{GeV}^{-2}]$ &             $[1,10]$ &     Proton size \\
        $B_{G,q}[\mathrm{GeV}^{-2}]$ &         $[0.05,3]$ &   Hot spot size \\
        $\mathrm{nCQP}$ &                      $[0,10]$ &    Number of constituent quarks \\
        $\sigma$ &                             $[0,1.5]$ &   Magnitude of $Q_s$ fluctuations \\
        $Q_s/(g^2\mu)$ &                       $[0.05,1.5]$ & Ratio of color charge density and saturation scale \\
        $m_\mathrm{JIMWLK}$ &                  $[0.2,2]$ &   Infrared regulator \\
        $\Lambda_\mathrm{QCD}[\mathrm{GeV}]$ & $[0,0.28]$ &   Spatial $\Lambda_\mathrm{QCD}$ \\
        $\mathrm{UVdamp}$ &                    $[0,1]$ &     Damping for high frequency models \\
        $\omega$ &                             $[0,10]$ &    Modification of proton shape \\
        \bottomrule
        \end{tabular}
        }
        \captionof{table}{Input parameters for the IP-Glasma particle collision simulator, along with their ranges and descriptions.}
        \label{table:qgp_params}
    \end{minipage}
    \hfill
    \begin{minipage}{0.45\linewidth}
        \centering
        \includegraphics[width=\linewidth]{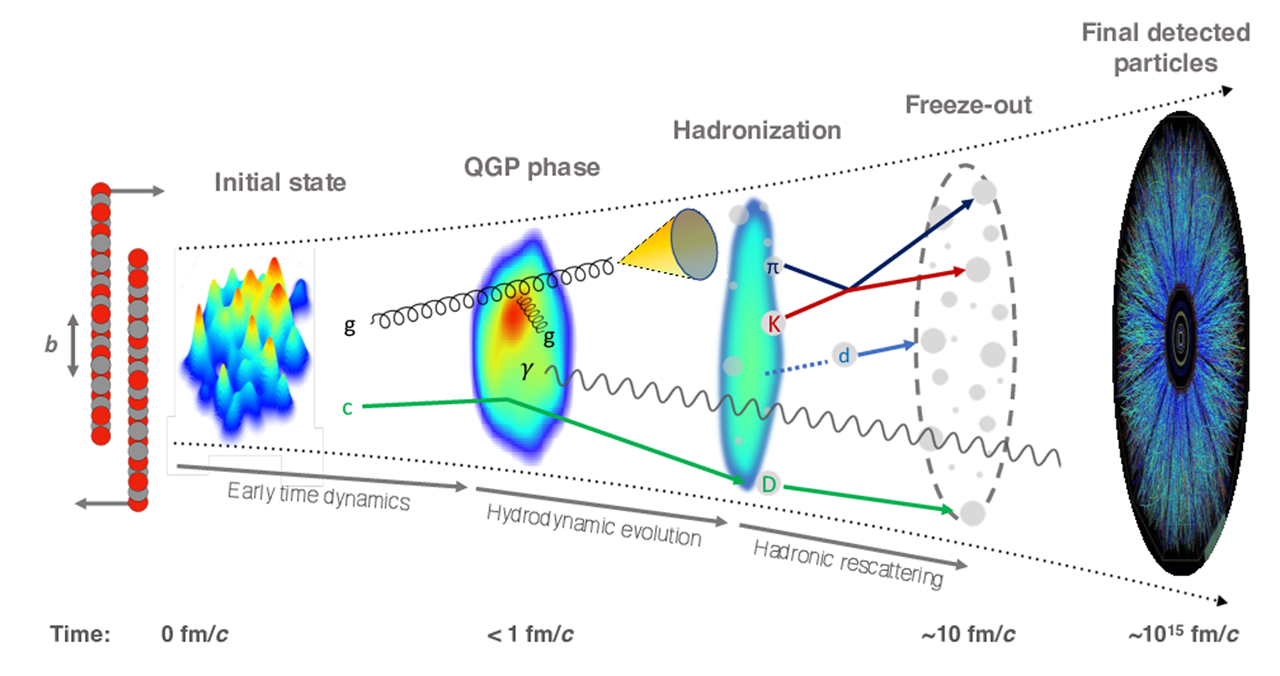}
        \captionof{figure}{Schematic illustration of a virtual particle collider simulator for generating the QGP. Reproduced from \cite{arslandok2023hot} with permission.}
        \label{fig:qgp}
    \end{minipage}
    \vspace{-.5cm}
\end{figure}

To investigate properties of this plasma, one further requires a virtual simulator that replicates the particle collision evolution for different choices of plasma parameters. \Cref{fig:qgp} visualizes each stage of this virtual particle collider: nuclei are first smashed together, the QGP is formed and evolved using relativistic hydrodynamic models, and the nuclear fluid is then converted to particles as it rapidly cools. Each run of this computer simulator is costly, requiring thousands of CPU hours to perform on supercomputing systems \cite{ji2021graphical}. Surrogate models have been used for emulating this expensive computer simulator \cite{ji2022multi,li2023additive}, leading to novel discoveries on the QGP \cite{ehlers2025bayesian,ehlers2026varp}. To ensure reliable scientific findings, recent QGP studies (see \cite{everett2021phenomenological}) make use of a collection of different virtual simulators for modeling particle collisions. In this case, different surrogate models need to be trained for these different (but related) simulators, which serve as different ``tasks'' in our multitask framework.

In what follows, we consider the ``IP-Glasma'' particle collision simulator proposed in \cite{ryu2015importance,ryu2018effects}, which adopts the IP-Sat model for the nucleon wavefunction and the classical Yang-Mills hydrodynamic evolution of the gluon fields. The selected observable (i.e., response) is the production rate of the $J/\psi$ meson at an energy level of 391.2 GeV, which provides insight into the underlying collision dynamics and quantum chromodynamics mechanisms. Here, we consider $T=3$ simulators (i.e., tasks) for surrogate modeling, with fixed infrared regulator level of $0.5$ GeV, $1.0$ GeV, and $1.5$ GeV, respectively. Such levels provide a reasonable range that covers experimental measurements from the ATLAS collaboration. All three simulators share the same $d=9$ input variables, whose ranges and descriptions are summarized in \Cref{table:qgp_params}. The goal is to construct accurate surrogate models for the $T=3$ related simulators over this nine-dimensional input space.

Our experimental setup is as follows. The input space from \Cref{table:qgp_params} is first normalized to the unit hypercube $[0,1]^9$. For each simulator (task), an equal number of $n=2^{10}=1024$ design points are used, with such design points sampled from randomly-shifted digital sequences satisfying \Cref{def:dseqs}. For the proposed fast MTGP, this is coupled with the digitally-shift-invariant kernels, satisfying the DSI-DSEQ variant in \Cref{cond:FMTGP}. As before, the lattice shift-variant (SI-LAT) variant in \Cref{cond:FMTGP} is not tested, since the periodicity requirement likely does not hold for the simulator here. 
The same testing points are used for each of the $T=3$ tasks. 

\begin{figure}[!t]
    \centering 
     \includegraphics[width=1\textwidth]{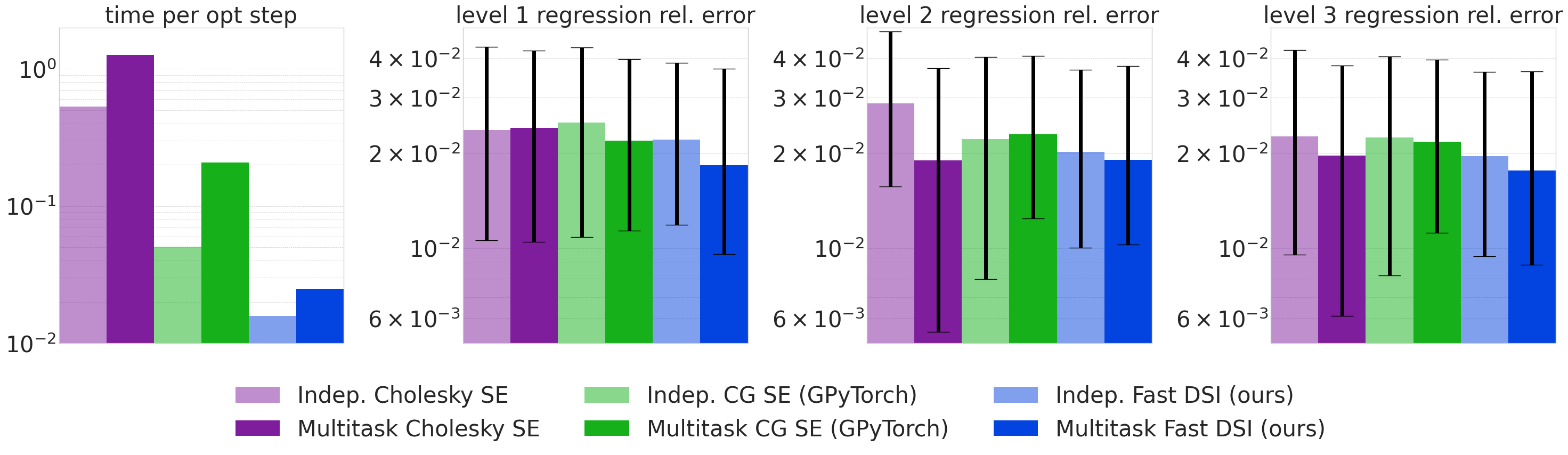}
    \caption{Comparison of GP fitting time per optimization step (seconds) and error for each task across various single-task and multitask GPs applied to the QGP problem. Bar heights are medians and the error bars mark 25-75 quantiles.}
    \label{fig:qgp_bar_charts}
    \vspace{-.9cm}
\end{figure}

\Cref{fig:qgp_bar_charts} shows the computation time for GP fitting (per optimization step) for each method, along with its test error for predicting each of the three particle collision simulators (i.e., tasks). As before, we see that the proposed fast MTGP provides slightly better predictions over existing benchmarks for all three tasks. The proposed multitask model also yields lower errors over its independent (i.e., single-task) counterpart, which shows the effectiveness of multitask learning here. More importantly, the proposed MTGP achieves this slightly improved predictive performance with considerably reduced computation time compared to existing benchmarks, allowing for much quicker model fits using the large training dataset. Thus, with a careful pairing of digital sequence (i.e., low-discrepancy) design points with a DSI kernel, our approach enables computationally efficient and accurate multitask surrogate modeling for the QGP application.

\section{Conclusions and Future Work} \label{sec:conclusion}

This article generalized fast Gaussian process (GP) regression using structured low-discrepancy (LD) sequences and certaint matching (digitally-)shift invariant kernels to the multitask GP (MTGP) setting. Our fast MTGP methods accommodate lattice and digital sequences with different shifts and different sample sizes for each task. We developed a scalable open-source implementation in the \texttt{FastGP} Python package and ran numerous benchmark problems to demonstrate the performance of the proposed methods.

In this article we have only tested fast MTGPs with the DSI-DSEQ variant fit by optimizing the MLE. Future experiments should be done to compare against the SI-LAT variant in \Cref{cond:FMTGP} and the GCV loss in \eqref{eq:MTGP_MLE_GCV_loss}. Multitask extensions of grid points and product kernels is also a valuable avenue for future work. 

\section*{Acknowledgements}

FH, PR, and AS are supported by NSF DMS 2316011. AS is partially supported by DARPA The Right Space HR0011-25-9-0031. YC and SM are supported by NSF DMS 2316012, NSF CSSI 2514008 and DE-AC02-05CH11231.


AS acknowledges that this material is based upon work supported by the U.S. Department of Energy, Office of Science, Office of Work-force Development for Teachers and Scientists, Office of Science Graduate Student Research (SCGSR) program. The SCGSR program is administered by the Oak Ridge Institute for Science and Education for the DOE under contract number DE-SC0014664.

PR and AS acknowledge that this work was supported by the Laboratory Directed Research and Development (LDRD) program and the Advanced Simulation and Computing, Verification and Validation (ASC V\&V) program at Sandia National Laboratories.

PR and AS acknowledge that this article has been co-authored by employees of National Technology and Engineering Solutions of Sandia, LLC under Contract No.\,DE-NA0003525 with the U.S. Department of Energy (DOE). The employees co-own right, title and interest in and to the article and are responsible for its contents. The United States Government retains and the publisher, by accepting the article for publication, acknowledges that the United States Government retains a non-exclusive, paid-up, irrevocable, world-wide license to publish or reproduce the published form of this article or allow others to do so, for United States Government purposes. The DOE will provide public access to these results of federally sponsored research in accordance with the DOE Public Access Plan available at \url{https://www.energy.gov/downloads/doe-public-access-plan}. 

\bibliographystyle{siamplain}
\bibliography{main}

\end{document}